\documentclass[manuscript,revised,noblind]{geophysics}
\usepackage{natbib}
\usepackage{amssymb}
\usepackage{amsmath}
\usepackage{mathrsfs}
\usepackage{multirow}
\usepackage{booktabs}
\usepackage{algorithm}
\usepackage{algpseudocode}
\usepackage{amsmath}
\usepackage[colorlinks,linkcolor=blue,citecolor=blue]{hyperref}

\usepackage{graphicx}

\begin{document}

\title{Coarse-to-fine multi-resolution hash encoding for implicit full waveform inversion}

\address{
\footnotemark[1]School of Sciences, Southwest Petroleum University, Chengdu 610500, China. E-mail: 13668229412@163.com; minfan@swpu.edu.cn. \\
\footnotemark[2]King Abdullah University of Science and Technology, Thuwal 23955-6900, Kingdom of Saudi Arabia. E-mail: shaowen.wang@kaust.edu.sa (corresponding author); tariq.alkhalifah@kaust.edu.sa.}
\author{Linrong Wang\footnotemark[1], Shaowen Wang\footnotemark[2], Fan Min\footnotemark[1], and Tariq Alkhalifah\footnotemark[2]}
\lefthead{Wang et al.}
\righthead{Coarse-to-fine implicit FWI}

\renewcommand{\thefootnote}{\fnsymbol{footnote}}

\renewcommand{\figdir}{.} 

\maketitle

\begin{abstract}
Implicit full waveform inversion (IFWI) reparameterizes the velocity model with a neural network, but it suffers from spectral bias and requires many iterations to converge.
Multi-resolution hash encoding uses a learnable feature grid to recover high wavenumber details at multiscales and provides strong local representations.
However, activating all resolution levels from the start of the inversion introduces high wavenumber components, which can exacerbate cycle skipping.
In this paper, we propose a coarse-to-fine IFWI (C2F-IFWI) that activates the multi-resolution hash levels progressively rather than all at once.
Specifically, only a few coarse base levels are kept active at initialization, while the remaining levels are introduced progressively.
A frozen hash level contributes no gradient until it is activated, so the coarse levels resolve the long-wavelength background before the finer levels are introduced regardless of the frequency range of the data.
A continuous activation index increases with iterations and determines how many levels are active at any point in training.
Each newly admitted level is then faded in gradually through a smooth weighting rather than switched on abruptly, which keeps the optimization stable during transition.
This scheme allows model capacity to grow synchronously with wavenumber without adding any additional trainable parameters and with negligible computational overhead.
Experiments on the synthetic Overthrust and BP 2004 models and on the Viking marine field dataset show that, compared with the all-resolution hash baseline, the coarse-to-fine scheme converges faster, attains higher accuracy, and is markedly more robust to the choice of initial model and to strong data noise.
\end{abstract}

\section{Introduction}\label{section: introduction}
Full waveform inversion (FWI) is a high-resolution seismic imaging technique.
It estimates subsurface physical properties by minimizing the discrepancy between observed and simulated seismic wavefields \citep{tarantola1984inversion,virieux2009overview}.
FWI exploits the full information content of seismic recordings, including amplitude, phase, and traveltime.
Therefore, it has become one of the most powerful tools for velocity model building and subsurface characterization
\citep{pratt1999seismic,alkhalifah2016full,operto2013guided}.

Traditional FWI parameterizes the subsurface model on an explicit grid, assigning an independent model parameter to each grid point or cell.
This parameterization is physically intuitive and straightforward to implement, but at field scale it results in an inverse problem with millions of unknowns.
The resulting optimization is highly nonlinear and computationally demanding, and it is prone to cycle skipping resulting in entrapment in local minima \citep{symes2008migration,bunks1995multiscale}.
Numerous strategies have been proposed to mitigate these difficulties, including multiscale frequency continuation \citep{bunks1995multiscale}, extended-domain formulations \citep{vanleeuwen2013mitigating}, optimal-transport misfit functions \citep{Engquist2016OTFWI,Sun2019application}, and joint physics-and-data-driven inversion \citep{wu2026joint}.
Nonetheless, an explicit grid inherently ignores the spatial correlations among neighboring geological structures, which limits inversion robustness and typically calls for additional regularization.

Deep neural networks have recently shown strong capabilities for velocity modeling.
In particular, convolutional neural networks (CNNs) can predict velocity models directly from seismic observations \citep{araya2018deep,yang2019deep,wu2019inversionnet,zhang2019velocitygan}.
By learning complex data-to-model mappings, these methods reduce the online inversion cost once trained \citep{kazei2021mapping,mosser2020stochastic}.
However, supervised approaches require large labeled training sets.
Such data are difficult and expensive to obtain in geophysical practice.
More importantly, their accuracy degrades when the target differs from the training distribution.
This raises concerns about generalization and practical deployment \citep{wu2019inversionnet}.

Implicit neural representations (INRs) represent continuous signals and physical fields with coordinate-based neural networks \citep{sitzmann2020implicit,mildenhall2021nerf,park2019deepsdf}.
Unlike grid-based parameterizations, INRs map spatial coordinates directly to physical quantities.
This continuous formulation provides resolution independence, parameter efficiency, and smooth spatial regularization \citep{chen2021learning}.
Motivated by these properties, several works have introduced INRs into FWI, parameterizing the velocity model with a coordinate network in an approach referred to as implicit full waveform inversion (IFWI) \citep{Sun2023Implicit,sun2021theory,wang2025multiresolution,chen2025full,he2026m}.
The network weights are optimized through the standard FWI objective using a differentiable wave solver.
A plain coordinate network, however, suffers from spectral bias and learns the smooth background much faster than the sharp velocity contrasts that FWI must recover \citep{rahaman2019spectral}.
This limitation renders IFWI impractical, requiring more than $1000$ iterations to converge.

To counter this bias, positional encoding enriches the input coordinates before they enter the multilayer perceptrons (MLPs), so that the network can model higher-frequency signals more effectively \citep{vaswani2017attention,tancik2020fourier}.
Two families of encodings dominate current practice, Fourier feature encoding and multi-resolution hash encoding.
Fourier feature encoding maps the coordinate through sinusoidal basis functions at multiple frequencies, which broadens the neural tangent kernel spectrum and helps the MLP capture high-frequency variations \citep{tancik2020fourier,sitzmann2020implicit}.
However, its frequency is global and predefined, making it difficult to adapt to the heterogeneity and multi-scale characteristics of geological structures \citep{fathony2021multiplicative}.
Learning the Fourier frequencies during the inversion has been proposed to relax this rigidity \citep{kang2026implicit}.

In contrast to the global and predefined Fourier basis, multi-resolution hash encoding employs a hierarchy of learnable feature grids of increasing resolution \citep{mueller2022instant}.
Coordinates are mapped into compact hash tables through efficient indexing, and the retrieved features are interpolated across scales \citep{takikawa2021neural}.
The coarse levels vary slowly and capture smooth, large-scale trends, while the fine levels vary rapidly and capture sharp, local detail.
In this way, the encoding spans a wide range of wavenumbers with few parameters, and its adaptive multi-scale structure suits localized geological bodies of varying complexity \citep{mueller2022instant}.
However, the fine-resolution levels are difficult to optimize early in the inversion \citep{martel2021acorn}.
When all hash levels are active from the first iteration, the fine levels inject high-wavenumber updates before the corresponding frequencies are present in the data and before the long-wavelength background is in place.
These updates are unconstrained by the misfit, so they imprint grainy, spurious structure, destabilize the optimization, and aggravate cycle skipping.
The standard remedy in waveform inversion is multiscale frequency continuation, which reconstructs the model from low to high wavenumber \citep{bunks1995multiscale}.
The multi-resolution hash, with its explicit coarse-to-fine level hierarchy, is naturally suited to this strategy, yet conventional hash encoding ignores it and exposes all levels at once.

To overcome this difficulty, we propose C2F-IFWI, an implicit FWI framework that activates the multi-resolution hash levels from coarse to fine rather than all at once.
Only the coarsest levels are active at the start, and the remaining levels are unfrozen progressively as the inversion proceeds.
A frozen level contributes nothing to the encoding and receives no gradient, so its feature table is held at its initial value until the schedule admits it.
The coarse levels therefore establish the long-wavelength background first, and each finer level is introduced only after the coarser structure is in place, so the model is reconstructed from low to high wavenumber \citep{Tariq2018wavenumber}.
Deferring the fine levels in this way spares them from the poorly conditioned gradients that destabilize the optimization when they are activated too early \citep{martel2021acorn}, and it suppresses cycle skipping.
At each coordinate, the currently active hash features are mapped by a sine-activated MLP to a velocity perturbation that is added to a smooth background model.

The main contributions of this work are summarized as follows
\begin{itemize}
\item We propose C2F-IFWI, an implicit FWI method that reparameterizes the velocity model with a multi-resolution hash encoding and activates its resolution levels from coarse to fine rather than all at once.

\item The activation is controlled by a continuous index that advances with the iterations together with a smooth per-level weighting, so the coarse levels resolve the long-wavelength background before each finer level is faded in.
Since a frozen level has no gradient updates and the schedule adds no trainable parameters, the process reconstructs the model from low to high wavenumber and removes the high-wavenumber artifacts that arise when every level is active from the start.

\item Applied to the synthetic Overthrust and BP 2004 models, C2F-IFWI converges faster and to a lower model error than the all-resolution hash baseline, and it stays accurate as the initial model degrades from a smoothed model to a linear gradient and a homogeneous constant, where the baseline deteriorates sharply.
On the Viking marine field data, it produces a cleaner, well-tied velocity model with a lower data misfit than the baseline and a more structurally consistent image than conventional grid-based FWI.
\end{itemize}

\par
\vspace{0.4cm}

\section{Theory}\label{section: Preliminary}
Seismic wave propagation in a constant-density acoustic medium is governed by the scalar wave equation
\begin{equation}
\frac{1}{v(\mathbf{x})^{2}}\,\frac{\partial^{2} u(\mathbf{x},t)}{\partial t^{2}} - \nabla^{2} u(\mathbf{x},t) = s(\mathbf{x},t),
\label{eq:wave}
\end{equation}
where $\mathbf{x}\in\Omega\subset\mathbb{R}^{2}$ is the spatial coordinate in 2D, $v(\mathbf{x})$ the velocity model, $u(\mathbf{x},t)$ the pressure wavefield, and $s(\mathbf{x},t)$ the source term.
Sampling the wavefield at the receiver locations defines the nonlinear forward operator $\mathcal{F}$ that maps a velocity model to the predicted data, $\mathbf{d}_{\mathrm{syn}} = \mathcal{F}(v)$.
In practice the wave equation is solved with a time-domain finite-difference scheme that is made fully differentiable, so that gradients can propagate through the solver.
This differentiability is required for full waveform inversion (FWI), which computes the gradient of an objective function that measures, through a chosen norm, the misfit between the predicted data and the observed field data.

Implicit FWI (IFWI) is a form of FWI in which the velocity is represented with a coordinate-based neural network instead of free values on the simulation grid \citep{sun2021theory,Sun2023Implicit}.
The velocity at a coordinate $\mathbf{x}$ is
\begin{equation}
v_{\boldsymbol{\theta}}(\mathbf{x}) = g_{\boldsymbol{\theta}}\!\big(\gamma(\mathbf{x})\big),
\label{eq:ifwi_param}
\end{equation}
where $g_{\boldsymbol{\theta}}:\mathbb{R}^{c}\!\to\!\mathbb{R}$ is a network with weights $\boldsymbol{\theta}$ and $\gamma(\cdot)$ is an input encoding.
In this case, the inversion variable becomes the finite set of weights $\boldsymbol{\theta}$, estimated by minimizing the $\ell_{2}$ data misfit \citep{tarantola1984inversion,virieux2009overview},
\begin{equation}
\boldsymbol{\theta}^{*} = \arg\min_{\boldsymbol{\theta}}\;  \frac{1}{2}\,\big\| \mathcal{F}(v_{\boldsymbol{\theta}}) - \mathbf{d}_{\mathrm{obs}} \big\|_{2}^{2}.
\label{eq:ifwi_obj}
\end{equation}
Because both the wave solver and the network are differentiable, the gradient is
\begin{equation}
\frac{\partial J}{\partial \boldsymbol{\theta}} = \int_{\Omega}\frac{\partial J}{\partial v(\mathbf{x})}\,\frac{\partial v_{\boldsymbol{\theta}}(\mathbf{x})}{\partial \boldsymbol{\theta}}\,\mathrm{d}\mathbf{x},
\label{eq:chain}
\end{equation}
where $\partial J/\partial v$ is the gradient with respect to the velocity and $\partial v_{\boldsymbol{\theta}}/\partial\boldsymbol{\theta}$ is obtained by back-propagation, the two computed by automatic differentiation \citep{griewank2008evaluating}.

Following SIREN \citep{sitzmann2020implicit}, $g_{\boldsymbol{\theta}}$ is an MLP with sinusoidal activations.
With encoded input $\mathbf{h}_{0}=\gamma(\mathbf{x})$
\begin{align}
&\mathbf{h}_{l} = \sin\!\big(\omega\,\mathbf{W}_{l}\,\mathbf{h}_{l-1}\big), \qquad l = 1,\dots,L, \label{eq:siren_hidden}\\
&g_{\boldsymbol{\theta}}(\mathbf{x}) = \mathbf{W}_{L+1}\,\mathbf{h}_{L}, \label{eq:siren_out}
\end{align}
where $\{\mathbf{W}_{l}\}$ are the bias-free weights and $\omega$ scales the pre-activations.
Setting $\gamma(\mathbf{x})=\mathbf{x}$ reduces to the coordinate network, which exhibits spectral bias, fitting low-frequency content much faster than high-frequency content \citep{rahaman2019spectral,tancik2020fourier}.
The encoding $\gamma$ is the lever that reduces this bias.

The encoding Eq.~\eqref{eq:ifwi_param} acts as a generic interface.
It is any map $\gamma:\mathbb{R}^{2}\!\to\!\mathbb{R}^{c}$ that transforms the raw coordinate into a $c$-dimensional feature vector before the MLP $g_{\boldsymbol{\theta}}$ acts on it.
A subsurface velocity model often contains both smooth background trends and sharp interfaces, so the high-frequency content that the plain network learns slowly is exactly the content that the inversion must recover.
To counter the bias, $\gamma$ must inject high-frequency structure into the input of $g_{\boldsymbol{\theta}}$ before the sinusoidal layers act on it.
We instantiate $\gamma$ as a multi-resolution hash encoding, a set of learnable local features defined over multiple resolutions.

The encoding we use is a multi-resolution hash encoding in the style of Instant-NGP \citep{mueller2022instant,takikawa2021neural}.
It arranges learnable features over $K$ resolution levels, where $K$ is independent of the SIREN depth $L$ in Eq.~\eqref{eq:siren_hidden}.
Level $k$ has side resolution
\begin{equation}
\begin{aligned}
N_{k} &= \big\lfloor N_{\min}\, b^{\,k-1} \big\rfloor, \qquad k = 1,\dots,K,\\
b &= \exp\!\left(\frac{\ln N_{\max} - \ln N_{\min}}{K-1}\right),
\end{aligned}
\label{eq:hash_res}
\end{equation}
so the resolutions grow geometrically from a coarse $N_{\min}$ to a fine $N_{\max}$, with $N_{1}=N_{\min}$ and $N_{K}=N_{\max}$.
Each level $k$ owns a learnable feature table $\boldsymbol{\theta}^{H}_{k}\in\mathbb{R}^{T\times F}$ with $T$ entries of feature dimension $F$.
To encode $\mathbf{x}$ at level $k$, the coordinate is scaled to that level by the factor $N_{k}$ and its enclosing grid cell is identified.
Each of the $2^{2}=4$ integer corners $\mathbf{c}$ of that cell is mapped to a table slot by the spatial hash
\begin{equation}
h(\mathbf{c}) = \left(\bigoplus_{i=1}^{2} c_{i}\,\pi_{i}\right) \bmod T,
\label{eq:hash_lookup}
\end{equation}
where $\bigoplus$ denotes the bitwise XOR and $\pi_{1},\pi_{2}$ are large fixed primes.
The $F$-dimensional features fetched at the four corners are bilinearly interpolated according to the position of $\mathbf{x}$ within the cell, yielding a per-level feature $\mathbf{f}_{k}(\mathbf{x})\in\mathbb{R}^{F}$ that is a continuous and almost everywhere differentiable function of $\mathbf{x}$.
Concatenating across levels gives the encoding
\begin{equation}
\gamma_{H}(\mathbf{x}) = \big[\mathbf{f}_{1}(\mathbf{x}),\dots,\mathbf{f}_{K}(\mathbf{x})\big] \in \mathbb{R}^{KF},
\label{eq:hash_concat}
\end{equation}
so the feature dimension is $c = KF$, which sets the input width of $g_{\boldsymbol{\theta}}$ through $\mathbf{h}_{0}=\gamma_{H}(\mathbf{x})$.
This encoding addresses the spectral bias directly.
The coarse levels with small $N_{k}$ vary slowly across $\Omega$ and capture smooth, global trends, while the fine levels with large $N_{k}$ vary rapidly and capture local detail.
Concatenating them gives $g_{\boldsymbol{\theta}}$ direct access to high-frequency content at the input, so the high-frequency detail is read out from the fine-level tables rather than being synthesized by the MLP alone \citep{mueller2022instant,takikawa2021neural}.
The table entries $\{\boldsymbol{\theta}^{H}_{k}\}$ are learnable and optimized jointly with the MLP weights, so the inversion variable $\boldsymbol{\theta}$ in Eqs.~\eqref{eq:ifwi_param}, \eqref{eq:ifwi_obj}, and \eqref{eq:chain} comprises both the network weights $\{\mathbf{W}_{l}\}$ and the hash tables $\{\boldsymbol{\theta}^{H}_{k}\}$, and the chain rule in Eq.~\eqref{eq:chain} back-propagates the data misfit into the tables as well.
The data therefore drive each level's local features during the inversion, so the representable detail is local and adapts spatially.

\begin{algorithm}[t]
\caption{C2F-IFWI}
\label{alg:hybridifwi}
\begin{algorithmic}[1]
\Require Observed data $\mathbf{d}_{\mathrm{obs}}$; Source wavelet at peak frequency
         $f$; Initial model $v_{0}$; Scale $\sigma$; epochs $E$; Learning rate $\eta$;
         Coarse levels $K_{0}$; Ramp fraction $\rho$
\Ensure  Inverted velocity $v_{\boldsymbol{\Theta}}$
\State Build normalized grid $\mathcal{X}=\{\mathbf{x}_{i}\}\subset[0,1]^{2}$.
\State Init INR backbone $\boldsymbol{\theta}_{g}$ and hash tables
       $\{\boldsymbol{\theta}^{H}_{k}\}_{k=1}^{K}$.
\State Init activation $w_{k}\!\gets\!1$ for $k\le K_{0}$, else $0$
\For{$e = 1$ to $E$}
  \State Update $\{w_{k}\}$ from the cosine ramp $\tau(e)$
         \Comment{Eq.~\eqref{eq:m_tau},\eqref{eq:m_wk}}
  \State $\mathbf{E}\!\gets\!
         \mathrm{diag}(\mathbf{w})\,\textsc{HashEncode}(\mathcal{X};\{\boldsymbol{\theta}^{H}_{k}\})$
         \Comment{Eq.~\eqref{eq:m_hash_c2f}}
  \State $V\!\gets\!v_{0}+\sigma\, g_{\boldsymbol{\theta}_{g}}(\mathbf{E})$
  \State $\mathbf{d}_{\mathrm{syn}}\!\gets\!\mathcal{F}_{f}(V)$
  \State $J\!\gets\!\tfrac{1}{2}\|\mathbf{d}_{\mathrm{syn}}-\mathbf{d}_{\mathrm{obs}}\|_{2}^{2}$
  \State back-propagate $\nabla_{\boldsymbol{\Theta}}J$
  \State Adam step on $\boldsymbol{\Theta}$ (rate $\eta$)
\EndFor
\State \Return $v_{\boldsymbol{\Theta}}$
\end{algorithmic}
\end{algorithm}

\section{Method}\label{section: Methods}

\begin{figure}[htbp]
  \centering
  \includegraphics[width=0.95\textwidth]{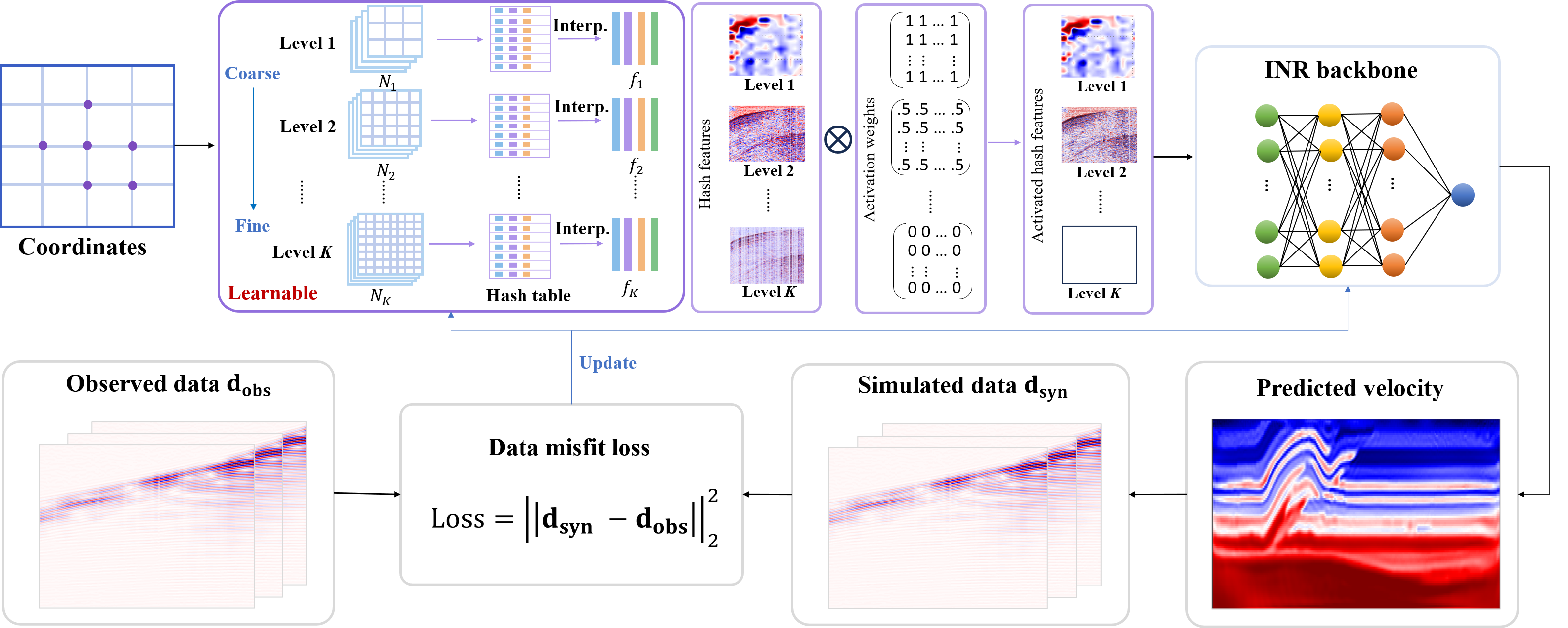}
  \caption{Overview of the proposed C2F-IFWI. Spatial coordinates are encoded by a multi-resolution hash whose levels are activated from coarse to fine over the inversion. The active features are decoded to the velocity by a sine-activated multilayer perceptron, and a differentiable acoustic solver maps the velocity to predicted data that are compared with the observations to update the hash tables and the network weights.}
  \label{fig:framework}
\end{figure}

Fig.~\ref{fig:framework} illustrates the pipeline of C2F-IFWI.
Given a normalized coordinate $\mathbf{x}$, the multi-resolution hash branch produces learnable features $\gamma_{\mathrm{H}}(\mathbf{x})$, whose resolution levels are unfrozen progressively from coarse to fine over the inversion.
The currently active hash features are passed to a sine-activated INR backbone, which outputs the predicted velocity.
A differentiable acoustic solver maps this velocity to simulated data $\mathbf{d}_{\mathrm{syn}}$, which are compared with the observed data $\mathbf{d}_{\mathrm{obs}}$ through the data-misfit loss.
This loss updates the hash tables and the backbone weights through backpropagation.

\subsection{Coarse-to-fine hash activation}
The hash branch returns the multi-resolution feature
$\gamma_{\mathrm{H}}(\mathbf{x})$ of Eq.~\eqref{eq:hash_concat}, obtained by
concatenating the interpolated features of $K$ learnable hash tables
$\{\boldsymbol{\theta}^{H}_{k}\}_{k=1}^{K}$.
The fine hash levels carry the highest-wavenumber content, but they are poorly conditioned early in the inversion.
When activated before the corresponding frequencies are present in the data, they inject high-wavenumber updates that the misfit cannot yet constrain.
These unconstrained updates destabilize training and aggravate cycle skipping.
Therefore, we do not activate all $K$ hash levels at once. We attach a per-level activation weight $w_{k}\in[0,1]$ to each level and mask the hash encoding level by level
\begin{equation}
\gamma_{\mathrm{H}}^{\mathrm{act}}(\mathbf{x})
= \big[\,w_{1}\mathbf{f}_{1}(\mathbf{x}),\;\dots,\;
        w_{K}\mathbf{f}_{K}(\mathbf{x})\,\big],
\label{eq:m_hash_c2f}
\end{equation}
where $\mathbf{f}_{k}(\mathbf{x})$ is the per-level feature of
Eq.~\eqref{eq:hash_concat}.
A level with $w_{k}=0$ contributes nothing to the encoding and, because the product $w_{k}\mathbf{f}_{k}$ blocks the gradient, receives no update, so its table is frozen until activated.
We keep the $K_{0}$ coarsest levels active throughout and unfreeze the remaining levels progressively, from coarse to fine, along a smooth schedule.
A continuous activation index
\begin{equation}
\tau(e) = K_{0} + \min\!\Big(1,\ \tfrac{e}{\rho E}\Big)\,(K-K_{0})
\label{eq:m_tau}
\end{equation}
grows from $K_{0}$ to $K$ over training, where $e$ is the current epoch, $E$ the total number of epochs, and $\rho$ the fraction of training after which all $K$ levels are active. 
The per-level weights track this index through a cosine ramp
\begin{equation}
w_{k}(e) =
\begin{cases}
1, & k \le \lfloor\tau(e)\rfloor,\\[2pt]
\tfrac{1}{2}\!\left[1-\cos\!\big(\pi\,(\tau(e)-\lfloor\tau(e)\rfloor)\big)\right],
   & k = \lfloor\tau(e)\rfloor+1,\\[2pt]
0, & k > \lfloor\tau(e)\rfloor+1,
\end{cases}
\label{eq:m_wk}
\end{equation}
so each finer level is admitted smoothly rather than switched on abruptly. 
At every epoch $e$, these weights $\{w_{k}(e)\}$ are inserted into the masked encoding of Eq.~\eqref{eq:m_hash_c2f}, so the activated features $\gamma_{\mathrm{H}}^{\mathrm{act}}(\mathbf{x})$ that are fed to the network expand from coarse to fine as the inversion proceeds.

\begin{figure}[htbp]
  \centering
  \includegraphics[width=0.95\textwidth]{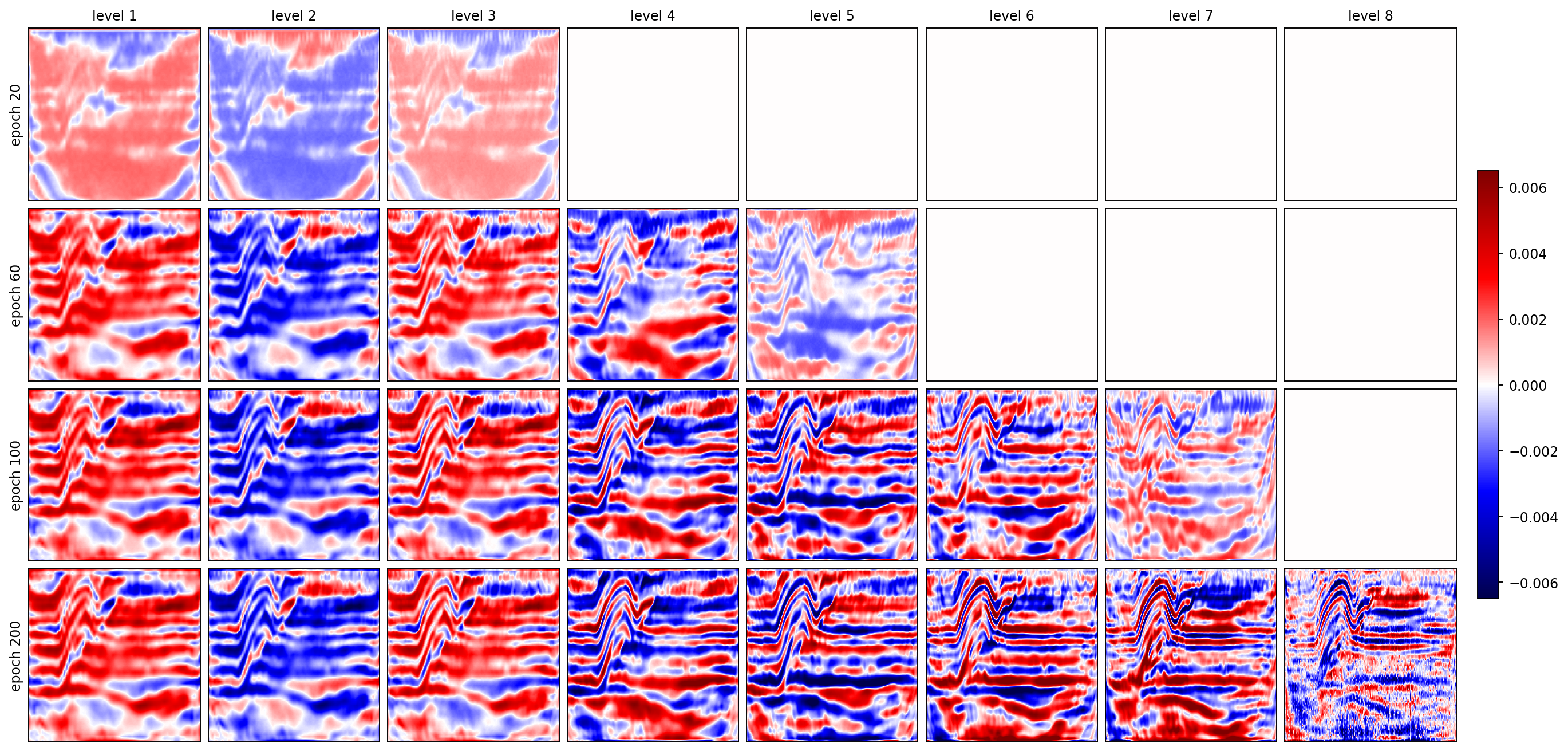}
  \caption{Coarse-to-fine use of the multi-resolution hash on Overthrust. Each panel shows one hash level's feature scaled by its activation weight (white means inactive, frozen at init). The rows are epochs 20/60/100/200 and the columns are levels 1 (coarsest) to 8 (finest), shown in a seismic colormap symmetric about zero. Only the coarse levels (1--3) are active early and capture the smooth background, while the coarse-to-fine schedule progressively unfreezes finer levels ($\approx$3$\rightarrow$5$\rightarrow$7$\rightarrow$8 active), which add high-wavenumber detail (faults, dipping reflectors) as the coarse levels keep the large-scale structure.}
  \label{fig:hash_usage_viz}
\end{figure}

Fig.~\ref{fig:hash_usage_viz} visualizes this behaviour on Overthrust.
Each panel shows one level's feature scaled by its activation weight, and the finer levels remain frozen (white) until the schedule unfreezes them, after which they add high-wavenumber detail while the coarse levels retain the large-scale structure.
By reconstructing the model from low to high wavenumber, this coarse-to-fine activation suppresses cycle skipping and stabilizes the early optimization of the fine levels.
For example, we set $K_{0}=2$, $K=8$, $\rho=0.6$, and $E=200$, so the schedule reaches full activation at epoch $\rho E=120$. 
At epoch $20$ (top row), $\tau=3$, so levels $1$ to $3$ are fully active with $w_{k}=1$ and levels $4$ to $8$ are frozen with $w_{k}=0$. 
At epoch $60$, $\tau=5$ and levels $1$ to $5$ are active. At epoch $100$, $\tau=7$ and only level $8$ is still frozen. 
From epoch $120$ onward, $\tau=8$ and all levels are active (bottom row). 
Between these snapshots the next level fades in gradually. 
For instance at epoch $50$, $\tau=4.5$ gives $w_{5}=0.5$ while levels $1$ to $4$ stay full.

\subsection{Velocity parameterization for IFWI}
Algorithm~\ref{alg:hybridifwi} summarizes the complete procedure.
At each coordinate, the coarse-to-fine activated hash features
$\gamma_{\mathrm{H}}^{\mathrm{act}}(\mathbf{x})$ of Eq.~\eqref{eq:m_hash_c2f} form the
encoding
\begin{equation}
\mathbf{e}(\mathbf{x})
= \gamma_{\mathrm{H}}^{\mathrm{act}}(\mathbf{x})\in\mathbb{R}^{\,KF},
\label{eq:m_concat}
\end{equation}
which is decoded by the sine-activated MLP backbone $g_{\boldsymbol{\theta}_{g}}$ of
Eqs.~\eqref{eq:siren_hidden}--\eqref{eq:siren_out}. Following the implicit
parameterization of Eq.~\eqref{eq:ifwi_param}, the predicted velocity is
\begin{equation}
v_{\boldsymbol{\Theta}}(\mathbf{x})
= v_{0}(\mathbf{x}) + \sigma\,
  g_{\boldsymbol{\theta}_{g}}\!\big(\mathbf{e}(\mathbf{x})\big),
\label{eq:m_velocity}
\end{equation}
where $v_{0}$ is the initial velocity model (often smooth) and $\sigma$ a fixed scaling constant.

The trainable parameters, namely the $K$ hash tables
$\{\boldsymbol{\theta}^{H}_{k}\}_{k=1}^{K}$ and the SIREN backbone weights
$\boldsymbol{\theta}_{g}$, are collected into a single set
$\boldsymbol{\Theta}=\boldsymbol{\theta}_{g}\cup\{\boldsymbol{\theta}^{H}_{k}\}_{k=1}^{K}$
and updated jointly by one Adam optimizer \citep{kingma2015adam} with a single learning
rate $\eta$. The per-level activation weights $\mathbf{w}=(w_{1},\dots,w_{K})$ are
advanced deterministically by the cosine ramp $\tau(e)$ rather than learned, so
gradient descent acts only on $\boldsymbol{\Theta}$.

Each epoch performs one forward pass and one update. The network forms the encoding
$\mathbf{e}(\mathbf{x})$ of Eq.~\eqref{eq:m_concat} from the coarse-to-fine activated
hash features and decodes it through the sine-MLP backbone into the velocity field
$v_{\boldsymbol{\Theta}}$ of Eq.~\eqref{eq:m_velocity}. The differentiable solver
$\mathcal{F}_{f}$ maps this velocity to synthetic data, which are compared with the
observed waveforms through the $\ell_{2}$ data misfit
\begin{equation}
J(\boldsymbol{\Theta})
= \,\big\| \mathcal{F}_{f}(v_{\boldsymbol{\Theta}})
     - \mathbf{d}_{\mathrm{obs}} \big\|_{2}^{2},
\label{eq:m_loss}
\end{equation}
and the gradient $\nabla_{\boldsymbol{\Theta}}J$ is back-propagated through the solver,
the backbone, and the active hash levels, after which the Adam step updates
$\boldsymbol{\Theta}$.

\section{Experiments}\label{section: Experiments}

\subsection{Experimental setup}
\label{subsec:exp_setup}
We evaluate C2F-IFWI on two synthetic benchmarks, the Overthrust and BP 2004 models, and on the Viking marine field dataset.
For the two synthetic models the observed data are generated from the known true model, whereas Viking consists of marine field recordings inverted with an estimated source wavelet.
The Viking data have a maximum offset of $3$~km, which is challenging.
All three cases use the same differentiable time-domain finite-difference solver for the constant-density acoustic wave equation.
The Overthrust model is a $94\times400$ grid at $25$~m spacing, rendering a $2.35\times10.0$~km$^2$ model recorded by $40$ shots at $250$~m spacing into $400$ fixed receivers and driven by a single $8$~Hz Ricker source.
The BP 2004 model is a $68\times188$ grid at $12.5$~m spacing, a $0.85\times2.35$~km$^2$ model recorded by $37$ shots at $62.5$~m spacing into $188$ fixed receivers and inverted over a multiscale frequency continuation with Ricker stages at $2$, $3$, $5$, and $8$~Hz.
The model used to invert the Viking dataset contains $320\times2281$ grid points at $12.5$~m spacing, a $4.0\times28.5$~km$^2$ model size, and the data were acquired using $1001$ shots at $25$~m spacing with a $120$-channel streamer and inverted over a multiscale sequence that widens the band from $2$--$6$~Hz to $2$--$20$~Hz.
All datasets share surface sources and receivers and an eighth-order spatial finite-difference stencil.
The remaining modeling and acquisition parameters are listed in Table~\ref{tbl:forward}.

\begin{table}[htbp]
\centering

\resizebox{0.85\textwidth}{!}{%
\begin{tabular}{lccc}
\toprule
Parameter & Overthrust & BP 2004 & Viking \\
\midrule
Data type                       & synthetic            & synthetic            & marine field \\
Grid ($n_{z}\times n_{x}$)      & $94\times400$        & $68\times188$        & $320\times2281$ \\
Model size (depth $\times$ length) & $2.35\times10.0$~km & $0.85\times2.35$~km & $4.0\times28.5$~km \\
Grid spacing                    & $25$~m               & $12.5$~m             & $12.5$~m \\
Time step $\Delta t$            & $2$~ms               & $1$~ms               & $4$~ms \\
Time samples $n_{t}$            & $3000$               & $8000$               & $1500$ \\
Record length                   & $6.0$~s              & $8.0$~s              & $6.0$~s \\
Source wavelet                  & Ricker, $8$~Hz       & Ricker, $2$--$8$~Hz  & estimated, $\sim$20~Hz \\
Frequency content               & single $8$~Hz        & multiscale, $2$--$8$~Hz & multiscale, $2$--$20$~Hz \\
Number of shots                 & $40$                 & $37$                 & $1001$ \\
Shot spacing                    & $250$~m              & $62.5$~m             & $25$~m \\
Receivers per shot              & $400$ (fixed)        & $188$ (fixed)        & $120$ (streamer) \\
Receiver spacing                & $25$~m               & $12.5$~m             & $25$~m \\
Source/receiver depth           & surface              & surface              & surface \\
Boundary                        & PML $20$             & PML $20$             & ABC $50$ + free surf. \\
FD spatial order                & $8$                  & $8$                  & $8$ \\
\bottomrule
\end{tabular}%
}
\caption{Forward-modeling and acquisition parameters. All datasets use the same differentiable time-domain finite-difference solver for the constant-density acoustic wave equation. Overthrust and BP 2004 are synthetic experiments, where the observed data are generated from the true model, while Viking is a marine field dataset whose observed data are field recordings inverted with an estimated source wavelet.}
\label{tbl:forward}
\end{table}

To probe robustness to the starting model, we invert each synthetic benchmark from three initial models of increasing difficulty, a smoothed version of the true model that retains the long-wavelength trend, a one-dimensional vertical gradient, and a homogeneous constant model.
The RMSE of these initial models to the truth rises from $418$ to $530$ to $1068$~m/s on Overthrust and from $549$ to $867$ to $1169$~m/s on BP 2004.

Table~\ref{tbl:configs} summarizes the configuration of the baseline and proposed methods.
Under identical data, initial model, backbone, and optimizer, we compare the IFWI baseline, which activates all hash levels from the first iteration, with the proposed C2F-IFWI, which activates them from coarse to fine.
For the Viking dataset, we additionally include a conventional grid-based FWI.
Both coordinate networks share a two-layer sine MLP backbone of width $128$ with $\omega=30$, together with a multi-resolution hash using a single feature per level, whose number of levels and resolution range are listed in Table~\ref{tbl:configs}.
The coarse-to-fine schedule keeps the two coarsest levels active and fades in the rest through a cosine ramp.
The two coordinate networks are trained with an Adam optimizer at learning rate $10^{-4}$, while the grid-based FWI uses a larger step size.
Overthrust is inverted for $200$ epochs from the smoothed initial model and for $1000$ epochs from the harder linear and constant initial models, BP 2004 for $500$ epochs, and Viking for $250$ epochs.
Overthrust and BP 2004 run on a single V100 and Viking on four V100s.

\begin{table}[htbp]
\centering

\resizebox{0.9\textwidth}{!}{%
\begin{tabular}{ll cccc ccc cc cc}
\toprule
 & & \multicolumn{4}{c}{Hash encoding} & \multicolumn{3}{c}{MLP backbone} & \multicolumn{2}{c}{Training} & \multicolumn{2}{c}{Cost} \\
\cmidrule(lr){3-6}\cmidrule(lr){7-9}\cmidrule(lr){10-11}\cmidrule(lr){12-13}
Dataset & Method & $L$ & base$\to$finest & $F$ & $\log_2 T$ & layers & width & $\omega$ & Epochs & LR & Params & Time \\
\midrule
\multirow{2}{*}{Overthrust}
 & IFWI     & $8$  & $64\!\to\!512$ & $1$ & $18$ & $2$ & $128$ & $30$ & $200$ & $10^{-4}$            & $0.62$M & $0.53$ \\
 & C2F-IFWI & $8$  & $64\!\to\!512$ & $1$ & $18$ & $2$ & $128$ & $30$ & $200$ & $10^{-4}$            & $0.62$M & $0.53$ \\
\midrule
\multirow{2}{*}{BP 2004}
 & IFWI     & $8$  & $64\!\to\!512$ & $1$ & $18$ & $2$ & $128$ & $30$ & $500$ & $10^{-4}$            & $0.62$M & $0.34$ \\
 & C2F-IFWI & $8$  & $64\!\to\!512$ & $1$ & $18$ & $2$ & $128$ & $30$ & $500$ & $10^{-4}$            & $0.62$M & $0.34$ \\
\midrule
\multirow{3}{*}{Viking}
 & FWI      & --   & --             & --  & --   & --  & --    & --   & $250$ & $5\!\times\!10^{-2}$ & $0.73$M & $7.09$ \\
 & IFWI     & $16$ & $16\!\to\!512$ & $1$ & $18$ & $2$ & $128$ & $30$ & $250$ & $10^{-4}$            & $0.75$M & $7.48$ \\
 & C2F-IFWI & $16$ & $16\!\to\!512$ & $1$ & $18$ & $2$ & $128$ & $30$ & $250$ & $10^{-4}$            & $0.75$M & $7.48$ \\
\bottomrule
\end{tabular}%
}
\caption{Configuration and cost on Overthrust, BP 2004, and Viking. $L$, $F$, and $T$ are the number of hash levels, features per level, and hash table size. Time is the measured wall-clock per epoch, including both forward modeling and inversion, in seconds, on one V100 for Overthrust and BP 2004 and on four V100s for Viking, so it is comparable only within each dataset. Parameters are in millions.}
\label{tbl:configs}
\end{table}

\subsection{The Overthrust model}
\label{subsec:exp_overthrust}
\begin{figure}[htbp]
  \centering
  \includegraphics[width=0.95\textwidth]{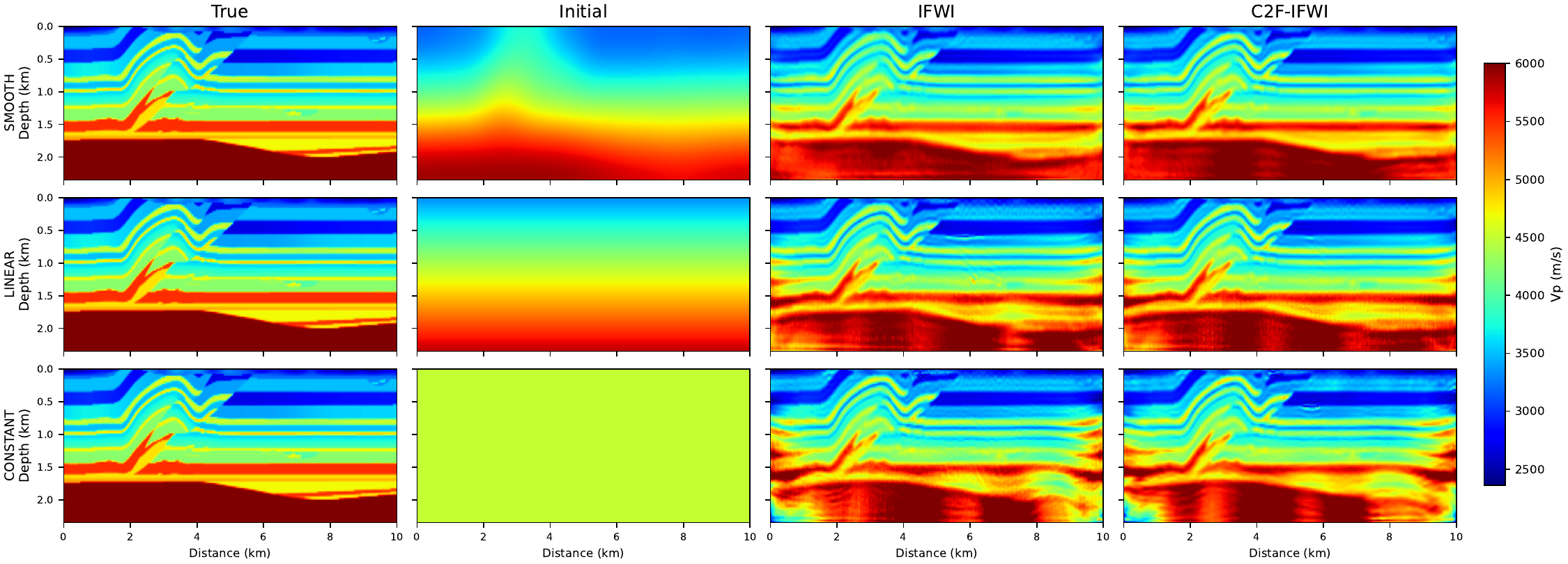}
  \caption{Inverted Overthrust velocity models from three initial models of increasing difficulty (rows, smoothed, linear, and constant). From left to right, the columns show the true model, the initial model, the IFWI result, and the proposed C2F-IFWI result.}
  \label{fig:overthrust}
\end{figure}

\begin{figure}[htbp]
  \centering
  \includegraphics[width=0.95\textwidth]{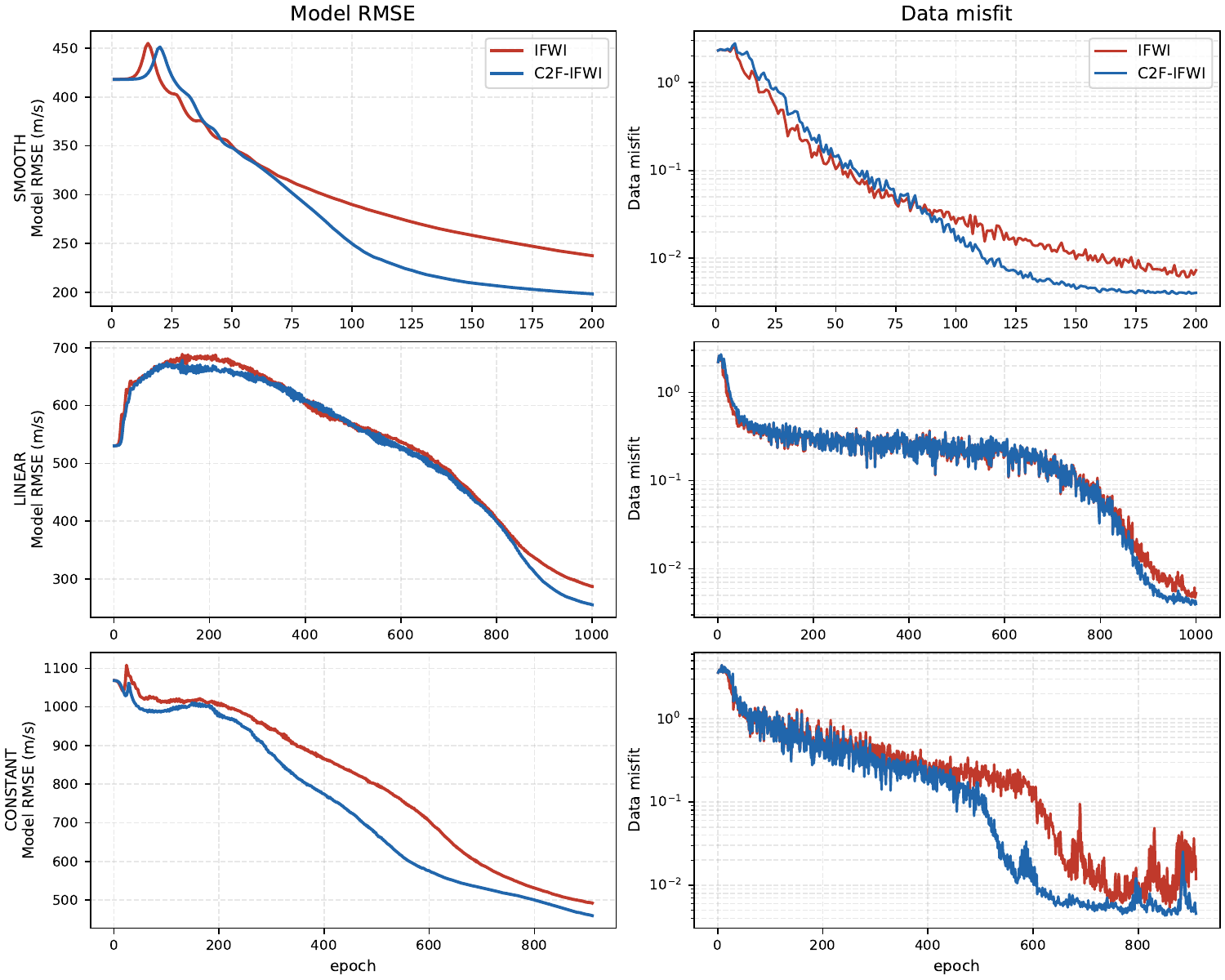}
  \caption{Convergence on Overthrust for the three initial models (rows, smoothed, linear, and constant). The left column shows the model RMSE against epoch and the right column the data misfit on a logarithmic scale. Red is the IFWI baseline with all resolution levels active and blue is the proposed C2F-IFWI.}
  \label{fig:overthrust_index}
\end{figure}

Fig.~\ref{fig:overthrust} shows the inverted Overthrust models for the three initial models (rows) and Fig.~\ref{fig:overthrust_index} the corresponding convergence, with the final model RMSE listed in Table~\ref{tbl:robust}.
The true model (first column) is a layered sedimentary sequence cut by a central thrust fault with dipping reflectors and velocity that increases with depth.
The initial models (second column) retain progressively less of this structure, from the smoothed long-wavelength trend, through a purely vertical gradient, to a homogeneous background that keeps nothing beyond the fixed near-surface.

\begin{table}[htbp]
\centering

\setlength{\tabcolsep}{20pt}
\begin{tabular}{l cc cc}
\toprule
 & \multicolumn{2}{c}{Overthrust} & \multicolumn{2}{c}{BP 2004} \\
\cmidrule(lr){2-3}\cmidrule(lr){4-5}
Initial model & IFWI & C2F-IFWI & IFWI & C2F-IFWI \\
\midrule
Smoothed & $237.0$ & $\mathbf{196.3}$ & $338.9$ & $\mathbf{259.8}$ \\
Linear   & $280.7$ & $\mathbf{250.1}$ & $410.9$ & $\mathbf{257.5}$ \\
Constant & $461.1$ & $\mathbf{437.0}$ & $512.4$ & $\mathbf{259.6}$ \\
\bottomrule
\end{tabular}
\caption{Final model RMSE (m/s) from three initial models of increasing difficulty on the two synthetic benchmarks. Lower is better, and the better of each IFWI and C2F-IFWI pair is in bold. C2F-IFWI improves on IFWI for every initialization, and on BP 2004 it stays almost invariant to the starting model while the baseline degrades sharply.}
\label{tbl:robust}
\end{table}

From every initial model the proposed C2F-IFWI (fourth column) is more accurate than the IFWI baseline (third column).
It lowers the model RMSE from $237.0$ to $196.3$~m/s from the smoothed start, from $280.7$ to $250.1$~m/s from the linear start, and from $461.1$ to $437.0$~m/s from the constant start (Table~\ref{tbl:robust}).
Because the IFWI baseline activates all hash levels from the first iteration, the fine levels imprint high-wavenumber updates before the long-wavelength background has settled, leaving a grainy roughness along the interfaces and in the deeper section that worsens as the initial model degrades.
C2F-IFWI instead admits the fine hash levels only after the coarse levels have resolved the background, so the faulted block, the dip of the reflectors, and the layering at depth are recovered crisply and continuously, and the model is largely free of the high-wavenumber texture seen in the IFWI result at every initialization.

The convergence curves (Fig.~\ref{fig:overthrust_index}) explain the gap.
For each initial model the left panel plots the model RMSE against epoch and the right panel the data misfit on a logarithmic scale.
Both networks bump above the initial-model level during the first few tens of epochs, when the early updates are still poorly constrained, and then descend.
The coarse-to-fine network decreases the model RMSE to the lower final value and reduces the data misfit faster, because it withholds the fine levels until the background is in place.
The IFWI baseline, which exposes all fine levels from the start, descends more slowly and plateaus at a higher RMSE, the residual high-wavenumber updates it admits early working against a clean fit.

\subsection{The BP 2004 model}
\label{subsec:exp_bp}
\begin{figure}[htbp]
  \centering
  \includegraphics[width=0.95\textwidth]{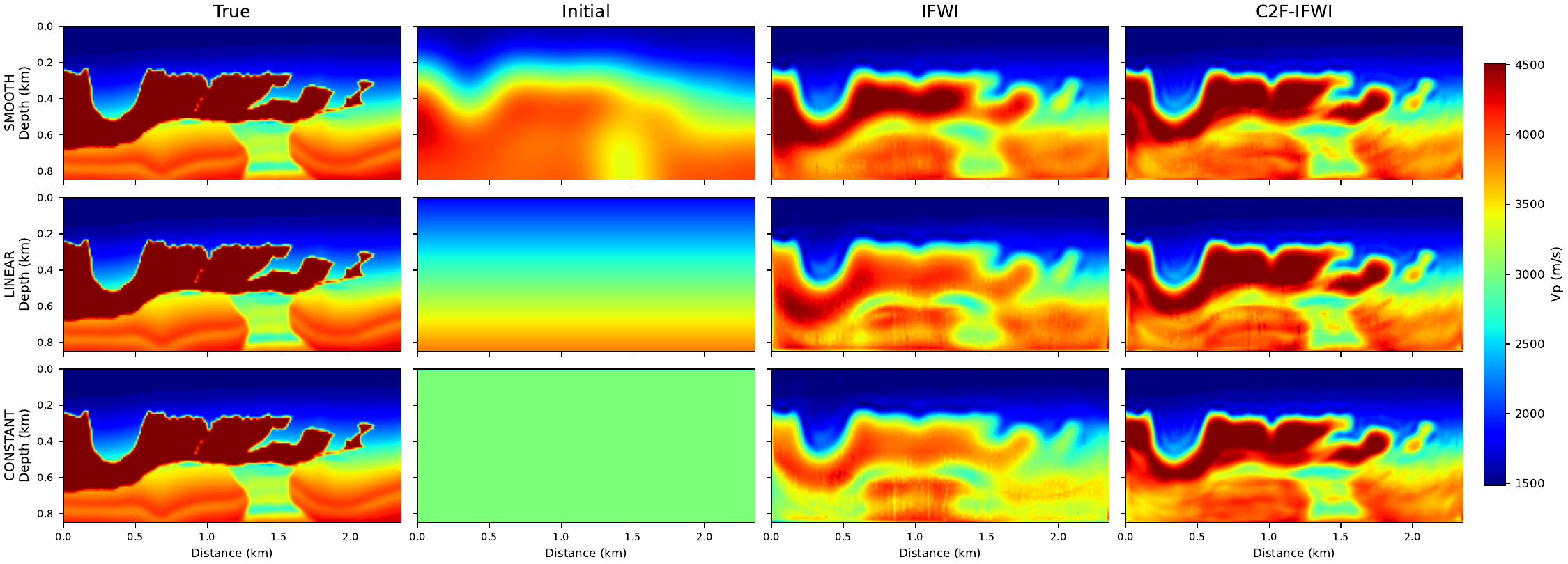}
  \caption{Inverted BP 2004 velocity models from three initial models of increasing difficulty (rows, smoothed, linear, and constant). From left to right, the columns show the true model, the initial model, the IFWI result, and the proposed C2F-IFWI result.}
  \label{fig:bp}
\end{figure}

\begin{figure}[htbp]
  \centering
  \includegraphics[width=0.95\textwidth]{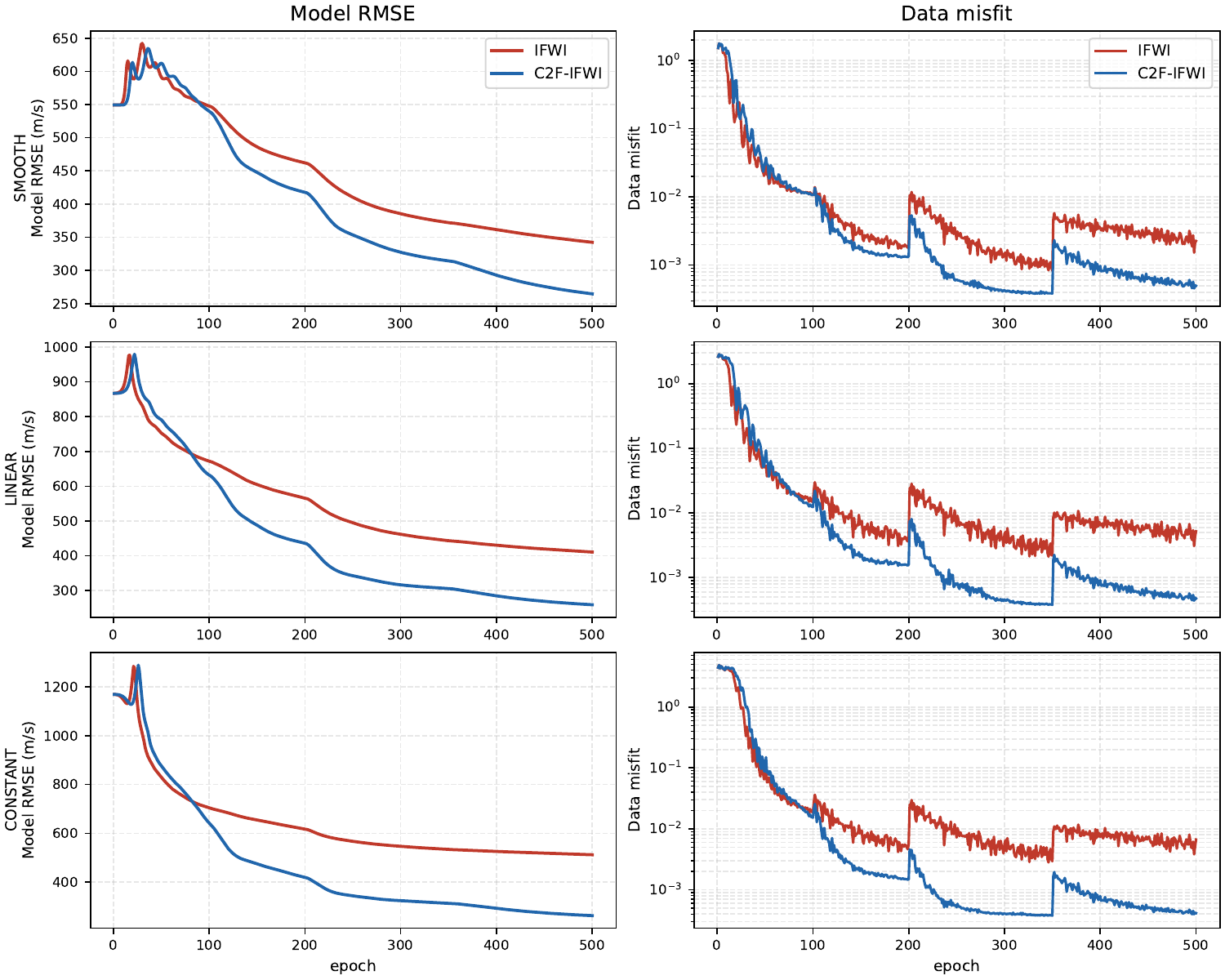}
  \caption{Convergence on BP 2004 for the three initial models (rows, smoothed, linear, and constant). The left column shows the model RMSE against epoch and the right column the data misfit on a logarithmic scale. The four multiscale stages (2, 3, 5, and 8~Hz) are visible as upward steps in the misfit. Red is the IFWI baseline with all resolution levels active and blue is the proposed C2F-IFWI.}
  \label{fig:bp_index}
\end{figure}

Fig.~\ref{fig:bp} shows the inverted BP 2004 models for the three initial models and Fig.~\ref{fig:bp_index} the corresponding convergence, again with the final RMSE in Table~\ref{tbl:robust}.
The true model (first column) contains a high-velocity salt body embedded in a layered sedimentary background, whose strong lateral contrasts make the reconstruction sensitive to the starting model.

The result confirms and sharpens the Overthrust finding.
The IFWI baseline degrades steadily as the initial model worsens, its RMSE rising from $338.9$~m/s for the smoothed start to $410.9$~m/s for the linear start and $512.4$~m/s for the constant start, because the always-active fine levels lock in high-wavenumber error that later iterations cannot undo.
C2F-IFWI is instead almost insensitive to the initialization, reaching $259.8$, $257.5$, and $259.6$~m/s from the three starts.
Deferring the fine levels lets the coarse levels first recover the long-wavelength background from whichever model the inversion starts, after which the fine detail settles onto interfaces that are already in place.
The improvement over IFWI therefore grows with the difficulty of the start, from $23\%$ for the smoothed model to $37\%$ and $49\%$ for the linear and constant models.

The convergence curves (Fig.~\ref{fig:bp_index}) show the same mechanism across the four multiscale stages.
Each frequency increase ($2$, $3$, $5$, and $8$~Hz) resets the data misfit upward, and within every stage C2F-IFWI reaches a lower misfit and a lower model RMSE than IFWI.
The separation is largest for the constant start, where the IFWI curve stalls at a high plateau while the C2F-IFWI curve continues to descend to essentially the same level it reaches from the smoothed start.
This near-invariance to the initial model is the clearest evidence that the coarse-to-fine schedule suppresses the initialization-dependent, cycle-skipping errors that limit the all-resolution baseline.

\subsection{Robustness to data noise}
\label{subsec:noise}

\begin{figure}[htbp]
  \centering
  \includegraphics[width=0.95\textwidth]{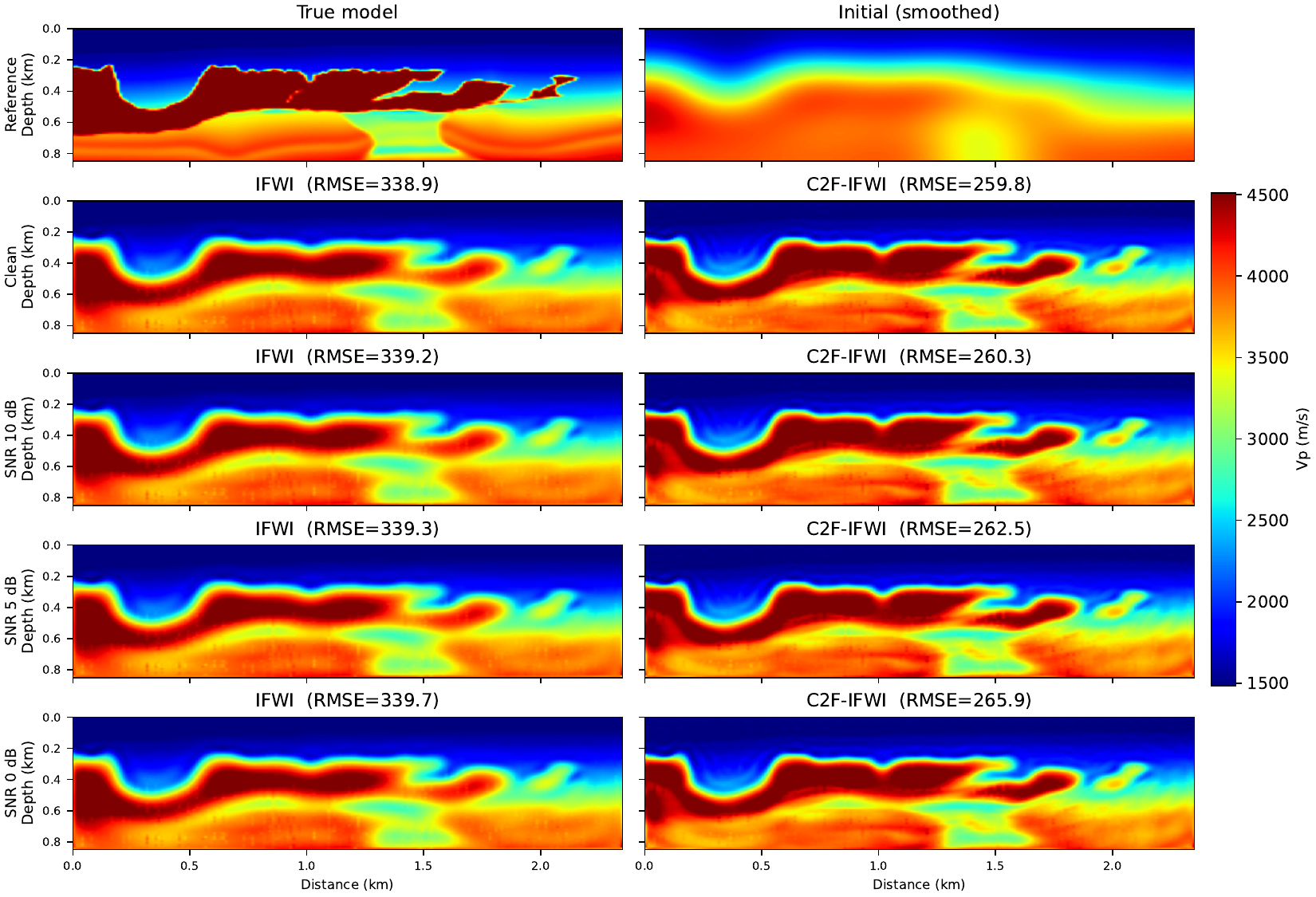}
  \caption{Inverted BP 2004 models under increasing data noise (rows, clean, SNR $10$, $5$, and $0$~dB) from the smoothed initial model. The top row shows the true and initial models, and the lower rows show the IFWI (left) and C2F-IFWI (right) reconstructions. The models are almost unchanged across the noise levels, and C2F-IFWI resolves the salt body and the central low-velocity channel more sharply than IFWI at every level.}
  \label{fig:noise}
\end{figure}

Because field records are never noise free, we test how the two methods behave when the observed data are contaminated.
We add zero-mean Gaussian noise to the synthetic Overthrust and BP 2004 records at three signal-to-noise ratios, $10$, $5$, and $0$~dB, with the noise standard deviation set from the data RMS, and invert from the smoothed initial model under the settings of the clean runs.
Fig.~\ref{fig:noise} shows the BP 2004 reconstructions, with the final model RMSE annotated in each panel.

\begin{figure}[htbp]
  \centering
  \includegraphics[width=0.95\textwidth]{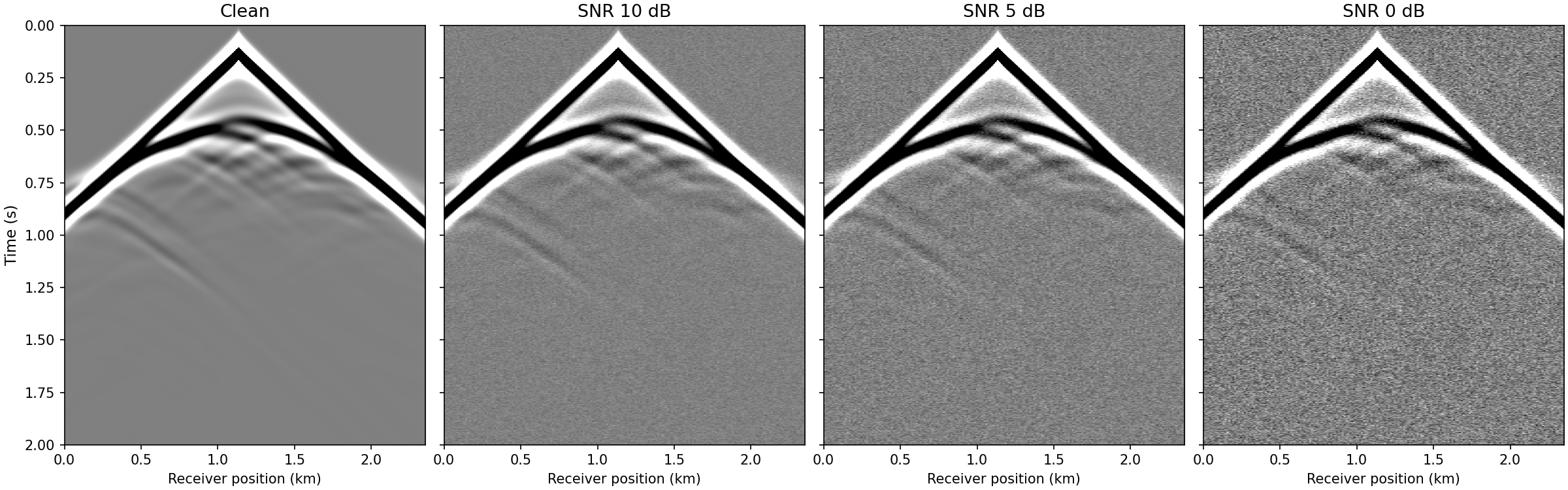}
  \caption{A representative BP 2004 observed shot gather (at $8$~Hz) at the four noise levels (clean, SNR $10$, $5$, and $0$~dB). The coherent first arrivals remain visible at every level, while the weaker late reflections are progressively buried and are almost lost at $0$~dB.}
  \label{fig:noise_obs}
\end{figure}

Both implicit methods are strongly robust to the added noise.
Even at $0$~dB, where the noise and the signal carry equal power, the model RMSE rises by only a few percent relative to the clean case, from $237.0$ to $242.7$~m/s for IFWI and from $196.3$ to $208.9$~m/s for C2F-IFWI on Overthrust, and by less than $1\%$ for IFWI on BP 2004.
The coordinate network cannot represent the incoherent high-wavenumber noise, so it acts as an implicit denoiser, and the coherent early arrivals that constrain the velocity survive even at $0$~dB, as Fig.~\ref{fig:noise_obs} shows for a representative gather where the strong first arrivals remain visible while the weaker late events are buried.
The BP 2004 reconstructions in Fig.~\ref{fig:noise} are correspondingly almost identical across the four noise levels.

The advantage of C2F-IFWI over the baseline persists at every noise level.
It stays ahead by $14$ to $17\%$ on Overthrust and by $22$ to $23\%$ on BP 2004, and the margin narrows only slightly as the noise grows.
The improvement is thus a property of the coarse-to-fine parameterization rather than of overfitting the clean data, and it carries over to the noisy regime that field data present.

\subsection{The Viking dataset}
\begin{figure}[htbp]
  \centering
  \includegraphics[width=0.95\textwidth]{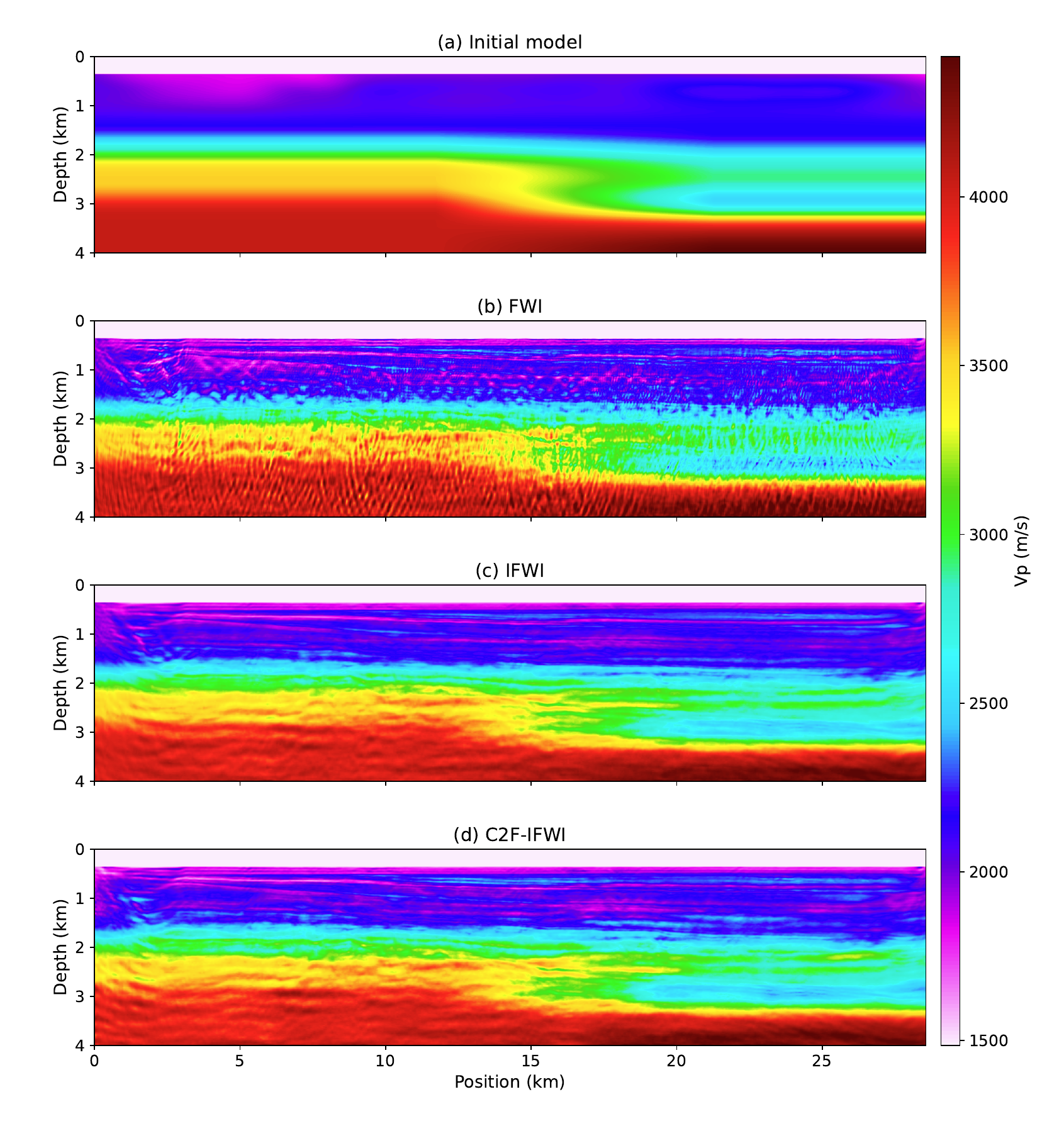}
  \caption{Inverted Viking velocity models. (a) Initial model, (b) grid-based FWI, (c) IFWI with all resolution levels active, and (d) the proposed C2F-IFWI.}
  \label{fig:viking_res}
\end{figure}

We further evaluate C2F-IFWI on the Viking 2D marine field dataset.
Fig.~\ref{fig:viking_res}(a) shows the initial model, obtained by interpolating the velocity profiles of two wells.
The traditional FWI result (Fig.~\ref{fig:viking_res}(b)) exhibits strong high-wavenumber oscillations and incoherent artifacts, more severe in the shallow section and toward the model edges.
These artifacts arise mainly from the limited illumination, the short-offset marine acquisition, and the ill-posed nature of gradient-based inversion when the model is parameterized directly on the grid.

Replacing the grid parameterization with an implicit coordinate network removes much of this noise.
The IFWI baseline (Fig.~\ref{fig:viking_res}(c)) recovers the main stratigraphic interfaces and is far cleaner than grid FWI, but because all resolution levels are active from the start, it reintroduces some irregular high-wavenumber features in the poorly illuminated shallow and edge regions.
The proposed C2F-IFWI (Fig.~\ref{fig:viking_res}(d)) removes this roughness and is the sharpest and cleanest reconstruction.
By deferring the fine hash levels until the long-wavelength background is in place, it lets the high-wavenumber detail settle onto interfaces that the lower frequencies have already constrained.
The thin layers and the laterally continuous events are therefore resolved more distinctly, and the layering stays continuous from the shallow section into the deeper part of the model, free of both the incoherent grid-FWI artifacts and the irregular features left by the IFWI baseline.

\begin{figure}[htbp]
  \centering
  \includegraphics[width=0.95\textwidth]{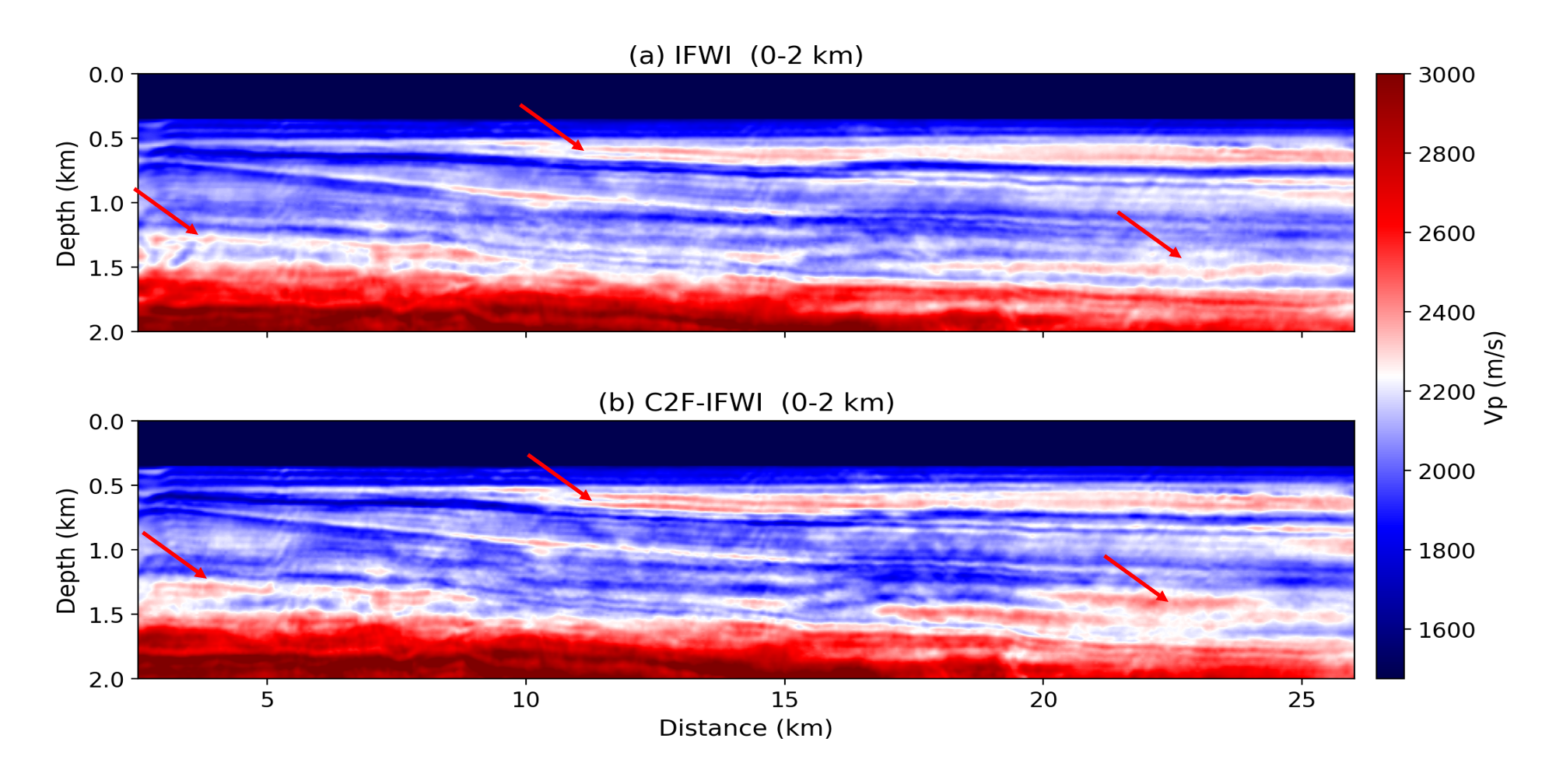}
  \caption{Magnified shallow $0$--$2$~km section of the inverted Viking models in a seismic colormap. (a) IFWI with all resolution levels active and (b) the proposed C2F-IFWI. Red arrows mark thin higher-velocity layers within the shallow low-velocity section that C2F-IFWI resolves more sharply and continuously than the baseline.}
  \label{fig:viking_sections}
\end{figure}

A magnified view of the shallow section makes this contrast explicit.
Fig.~\ref{fig:viking_sections} zooms into the upper $0$--$2$~km of the IFWI and C2F-IFWI models in a seismic colormap, where the thin layering dominates.
The IFWI section (Fig.~\ref{fig:viking_sections}(a)) recovers the main reflectors but leaves them slightly rougher and speckled with residual high-wavenumber noise.
The C2F-IFWI section (Fig.~\ref{fig:viking_sections}(b)) keeps the same reflectors laterally continuous and separates the thin layers more distinctly, confirming that the coarse-to-fine schedule sharpens the shallow structure rather than merely smoothing it.
The red arrows in both panels point to three thin higher-velocity layers embedded in the low-velocity shallow section, which the IFWI result renders as broken, speckled patches but C2F-IFWI recovers as continuous events.

\begin{figure}[htbp]
  \centering
  \includegraphics[width=0.95\textwidth]{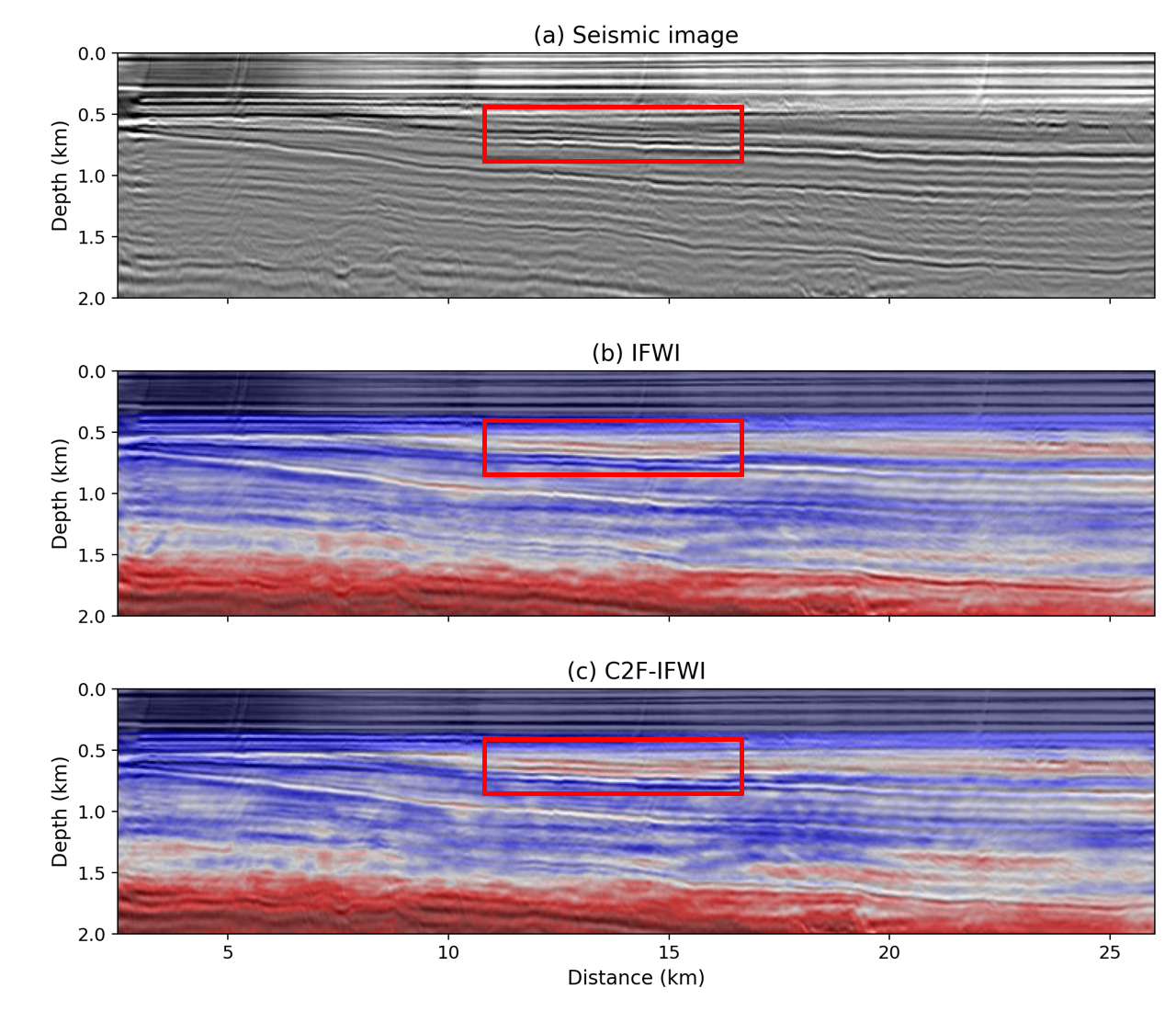}
  \caption{The overlap of the reverse-time-migration image and the inverted velocity models on the Viking field dataset. (a) The reverse-time-migration image obtained from the initial model. (b), (c) the same image overlaid semi-transparently on the (b) IFWI and (c) C2F-IFWI velocity models, both shown in a seismic colormap. The red box marks a zone of layered reflectors where the inverted velocity layering conforms to the migrated events, most continuously for C2F-IFWI. Migrated reflectors that align with the inverted velocity interfaces indicate a kinematically consistent model.}
  \label{fig:rtm}
\end{figure}

To further assess the geological consistency of the inverted models, we compare them with a reverse-time-migration image of the Viking data computed with the initial model (Fig.~\ref{fig:rtm}(a)), which provides an independent reference for the main reflector geometry and stratigraphic trends.
Figs.~\ref{fig:rtm}(b) and (c) overlay this image semi-transparently on the IFWI and C2F-IFWI velocity models.
For both implicit models the dominant velocity variations are broadly aligned with the reflector patterns of the image, and the shallow layered structures in the central section coincide with the migrated reflectors.
The red box marks one such zone, where the inverted velocity layering follows the migrated events.
The agreement is tightest for C2F-IFWI (Fig.~\ref{fig:rtm}(c)), whose sharper and more continuous interfaces align with the brightest migrated events inside the box, indicating that it recovers a geologically reasonable background while preserving the stratigraphic features.

\begin{figure}[htbp]
  \centering
  \includegraphics[width=0.95\textwidth]{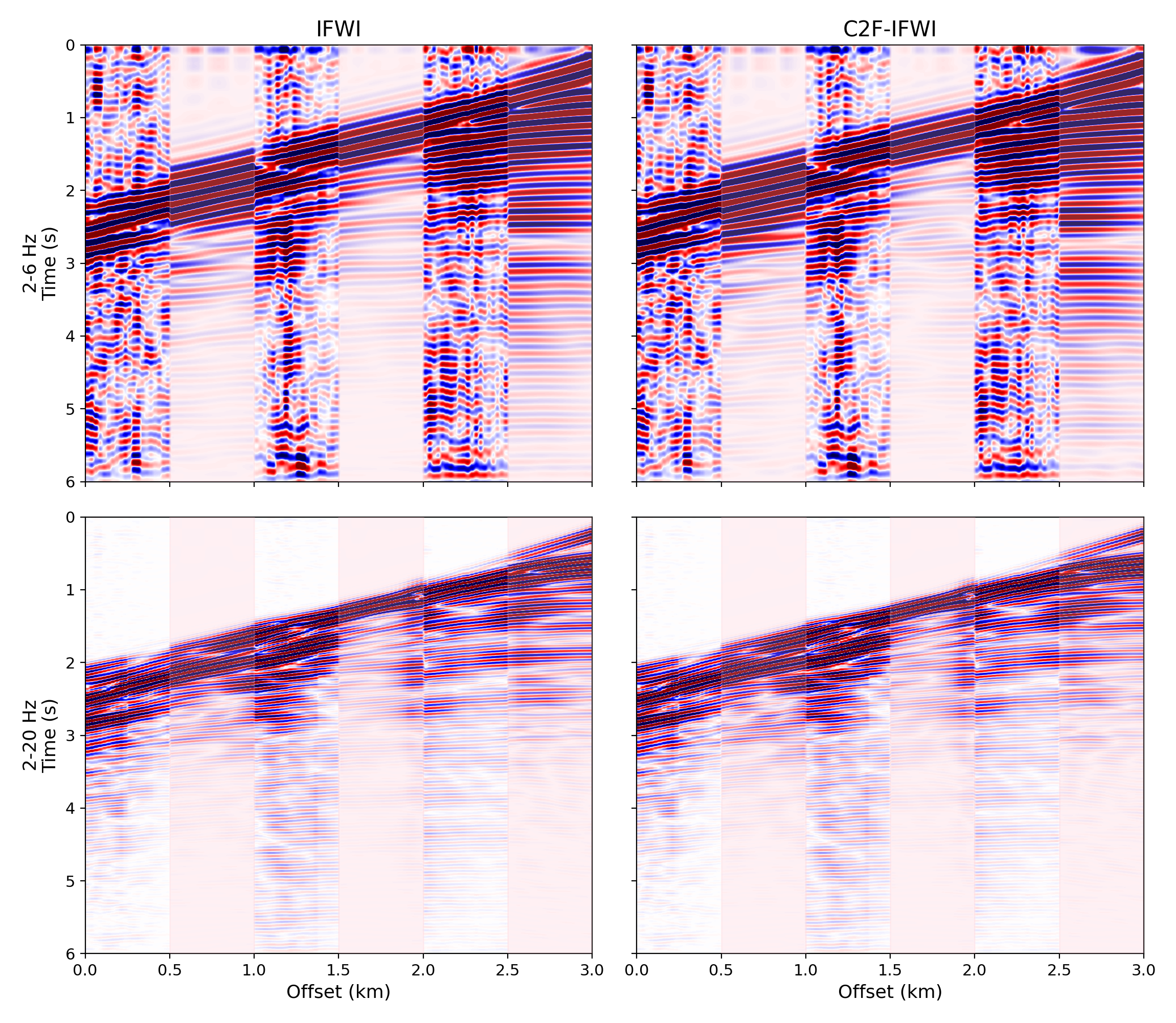}
  \caption{Observed versus synthetic data fit for a representative shot gather ($120$ traces over $0$--$3$~km offset) at two frequency bands, $2$--$6$~Hz (top) and $2$--$20$~Hz (bottom), for IFWI (left) and the proposed C2F-IFWI (right). The observed (obs) and synthetic (syn) data are interleaved in alternating blocks of $20$ traces, so an accurate fit appears seamless while a misfit shows as a discontinuity at the block boundaries.}
  \label{fig:obs_syn}
\end{figure}

We further assess the inversions directly in the data domain.
Fig.~\ref{fig:obs_syn} compares the observed records with the synthetic ones produced by forward modeling using the final inverted models, shown for a representative shot at the lowest and highest inversion bands ($2$--$6$ and $2$--$20$~Hz).
To allow the comparison to highlight small traveltime and phase differences, we interleave the observed and synthetic gathers in alternating blocks of twenty traces.
With this arrangement, an accurate fit appears as a single, seamless gather, while any misfit shows up as a discontinuity at the block boundaries.
At the long offsets both methods reproduce the recorded first arrivals and the main reflection packages, confirming that the inverted models are kinematically consistent with the data.
The IFWI synthetic events are slightly advanced relative to the adjacent observed events and leave a visible step at the splice, whereas the C2F-IFWI events cross the same boundaries more continuously.
The coarse-to-fine schedule therefore yields a closer data fit for these events, in line with the cleaner velocity model and the better-tied image it recovers (Figs.~\ref{fig:viking_res} and \ref{fig:rtm}).

\begin{figure}[htbp]
  \centering
  \includegraphics[width=0.95\textwidth]{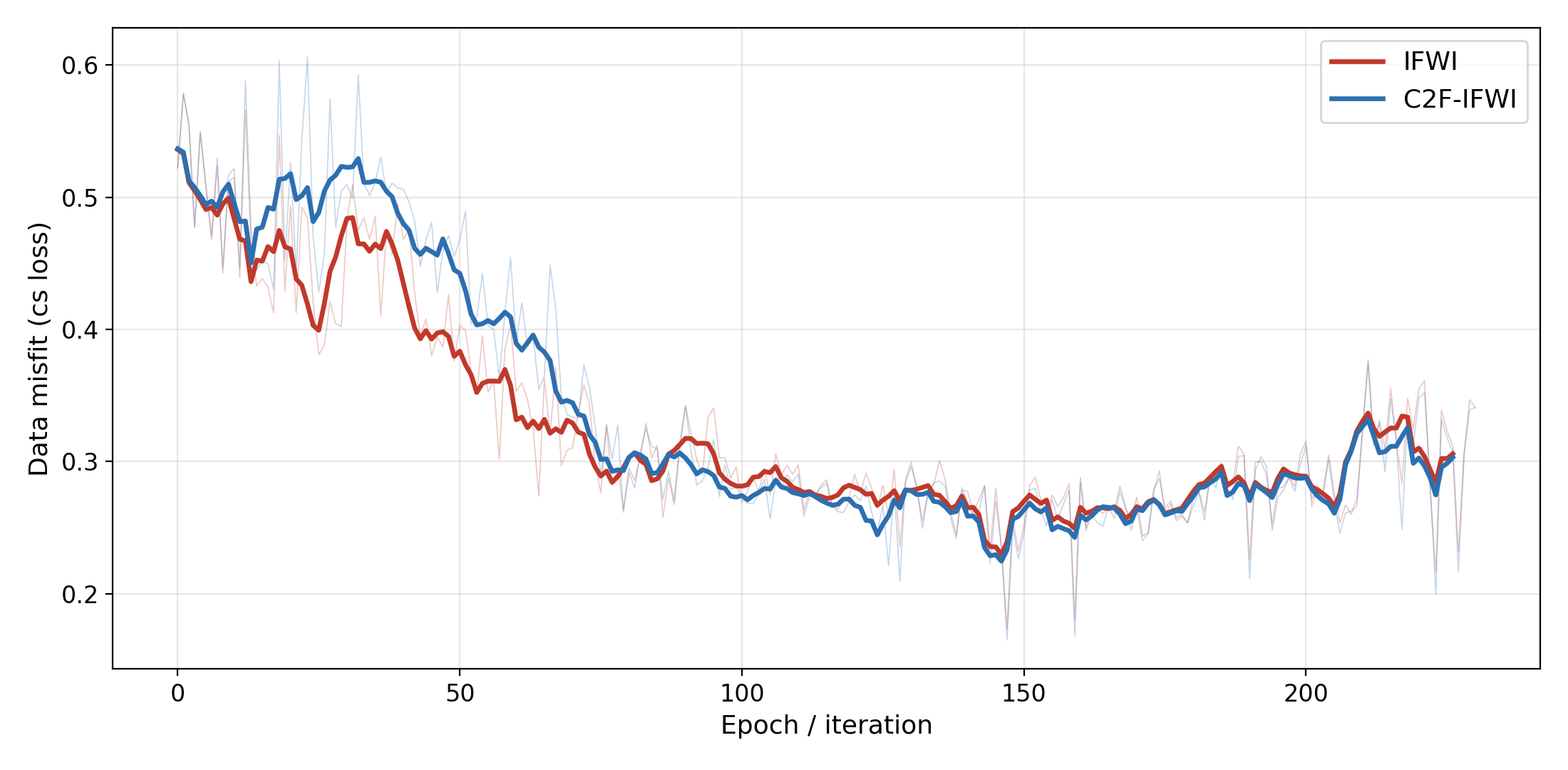}
  \caption{Data misfit on the Viking field data for the IFWI baseline (red) and the proposed C2F-IFWI (blue), both trained under identical settings.}
  \label{fig:viking_dataloss}
\end{figure}

Fig.~\ref{fig:viking_dataloss} shows the data misfit.
In the early stages of training, the data misfit of C2F-IFWI is consistently higher than that of IFWI.
This is an expected behavior of the scheduling scheme because the fine hash levels are frozen at this point, limiting the model's ability to fit the data.
Once the schedule has admitted all of the levels, near the middle of training, the two curves cross and C2F-IFWI continues to a lower misfit.

\begin{figure}[htbp]
  \centering
  \includegraphics[width=0.95\textwidth]{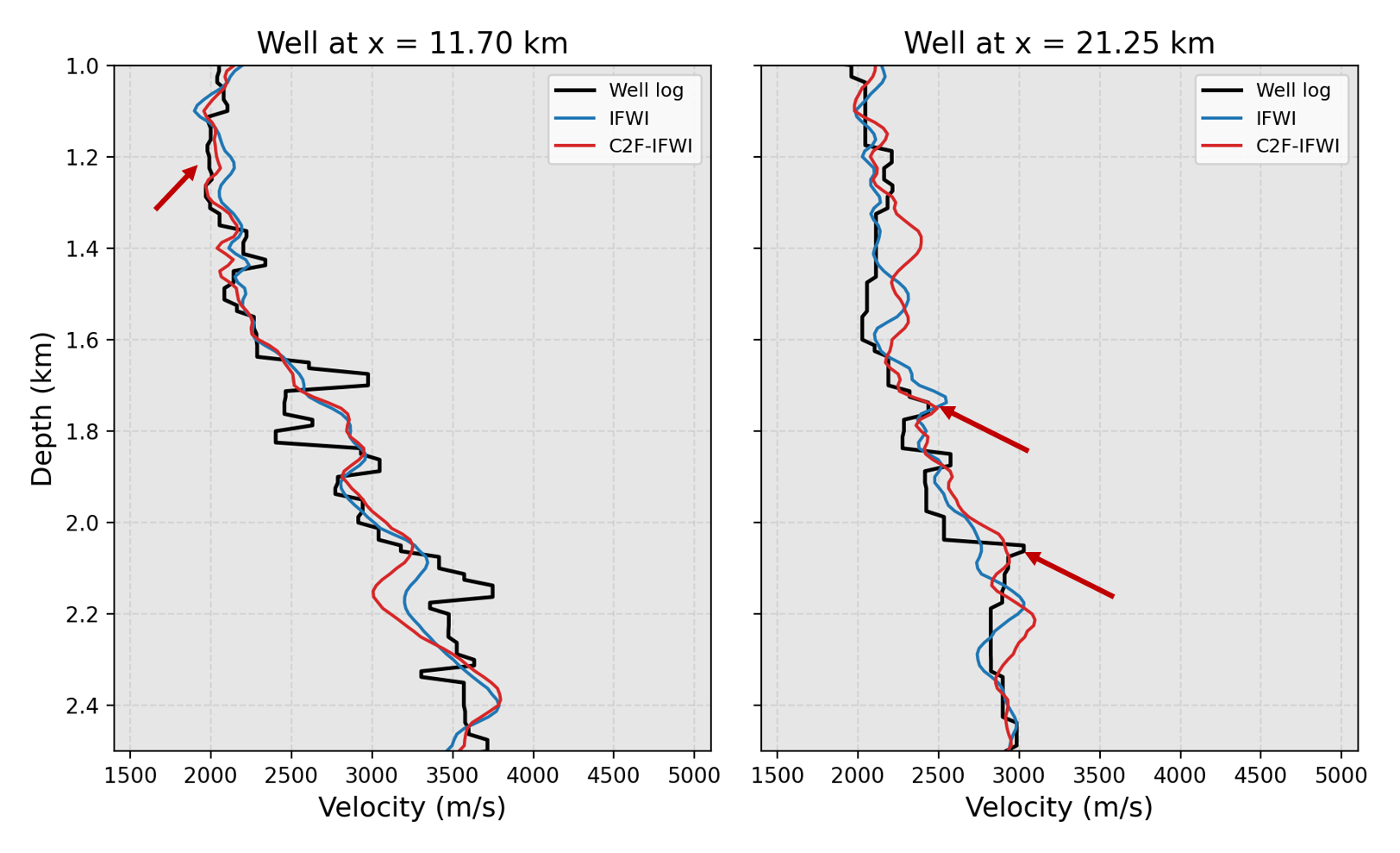}
  \caption{Velocity profiles at the two wells (Well-1 at $x=11.70$~km, Well-2 at $x=21.25$~km) over the $1.0$--$2.5$~km depth interval. Red arrows mark logged velocity steps where C2F-IFWI (red) gives the closest tie to the well log (black).}
  \label{fig:well}
\end{figure}

For a quantitative check, we extract vertical velocity profiles at the two well locations, Well-1 at $x=11.70$~km and Well-2 at $x=21.25$~km, over the $1.0$--$2.5$~km depth interval (Fig.~\ref{fig:well}).
At both wells the implicit methods follow the well-log velocity trends closely and comparably, capturing the major velocity increases with depth.
C2F-IFWI gives the closest and most consistent tie, tracking the logged velocity steps without the oscillations that contaminate the traditional FWI profile.
The red arrows point to such steps, near $1.2$~km at Well-1 and near $1.75$ and $2.1$~km at Well-2, where the C2F-IFWI profile follows the logged velocity kick that IFWI smooths through.
There are of course misties, and these are courtesy of anisotropy, which is a well-known issue causing depth misties \citep{alkhalifah2016research}.
Nevertheless, these close well ties confirm that the cleaner implicit models are quantitatively accurate rather than merely smoother.
Taken together, the model comparison, the migrated-image overlay, and the well ties show that C2F-IFWI transfers robustly to field data, where unconstrained grid FWI degrades.

\section{Discussion}\label{section: Discussion}
We first give a wavenumber-continuation analysis of why the coarse-to-fine activation helps, and then study each design choice empirically on the Overthrust model with the same settings as the experiments.

\subsection{Why coarse-to-fine activation helps}
\label{subsec:disc_theory}
We give a wavenumber-continuation argument that explains the faster convergence, the reduced roughness, and in particular the near-invariance of C2F-IFWI to the initial model.

The network writes the velocity as $v_{\boldsymbol{\theta}}(\mathbf{x}) = v_{0}(\mathbf{x}) + \sigma\, g_{\boldsymbol{\theta}_{g}}\!\big(\gamma_{\mathrm{H}}^{\mathrm{act}}(\mathbf{x})\big)$ with the masked features $\gamma_{\mathrm{H}}^{\mathrm{act}}$ of Eq.~\eqref{eq:m_hash_c2f}.
Hash level $k$ interpolates a grid of resolution $N_{k}$, so it can represent spatial wavenumbers only up to a grid Nyquist $\kappa_{k}$ proportional to $N_{k}$, which grows with $k$, and the weight $w_{k}$ gates how much of that band is admitted.

The data limit which of these wavenumbers can be constrained.
Linearizing the forward operator about the current model gives the Born relation between a model wavenumber and the recorded scattering.
For a source of angular frequency $\omega$ in a background velocity $v$, a scattering event with opening angle $\theta$ constrains the model wavenumber $|\mathbf{k}_{m}| = (2\omega/v)\cos(\theta/2)$, so the data reach at most $k_{\mathrm{data}}(\omega)=2\omega/v=4\pi f/v$ at backscattering \citep{sirgue2004efficient,Tariq2018wavenumber}.
Model wavenumbers above $k_{\mathrm{data}}(\omega_{\max})$ lie in the null space of the linearized map, and the misfit is, to first order, blind to them.

These two facts interact through the Jacobian $J=\partial \mathbf{d}_{\mathrm{syn}}/\partial\boldsymbol{\theta}$ and the Gauss--Newton Hessian $H=J^{\top}J$ \citep{pratt1999seismic,virieux2009overview}.
A level with $\kappa_{k}>k_{\mathrm{data}}(\omega_{\max})$ contributes columns of $J$ that are almost zero, so activating it from the first iteration appends vanishing eigenvalues to $H$ and worsens its conditioning, and it leaves the parameters of that level to be moved only by the higher-order and stochastic parts of the gradient.
Those updates surface as high-wavenumber artifacts and deepen the cycle-skipping minima of the nonconvex misfit \citep{bunks1995multiscale}.
This is the roughness and the initialization sensitivity that we observe in the IFWI baseline.

The coarse-to-fine schedule removes exactly these directions while they are uninformative.
Freezing a level sets $w_{k}=0$, which by Eq.~\eqref{eq:m_hash_c2f} drops its features and blocks its gradient, so the optimization is confined to the data-identifiable subspace where $H$ is better conditioned and the basin of attraction is wider.
The activation index of Eq.~\eqref{eq:m_tau} then unfreezes level $k$ only once the resolvable band has grown to reach $\kappa_{k}$, either because a multiscale stage has raised $k_{\mathrm{data}}(\omega)$, as on BP 2004 and Viking, or because the recovered coarse background has made the linearization accurate enough to constrain that wavenumber, as at the single frequency used on Overthrust.
The cosine weight of Eq.~\eqref{eq:m_wk} scales the corresponding columns of $J$ from zero to one, so the problem deforms continuously from the well-conditioned coarse subproblem to the full one.
This is a graduated-nonconvexity continuation carried out in model wavenumber, the parameter-space counterpart of the frequency continuation that multiscale FWI performs in the data \citep{bunks1995multiscale}, and because it does not rely on widening the data band it also helps at a single frequency.

The same argument predicts why the final accuracy is set by the data rather than by the start.
The coarse subproblem varies only the long-wavelength components that the low wavenumbers of the data constrain, so it is far less nonconvex and its minimizer barely depends on the fine-level initialization.
C2F-IFWI therefore recovers essentially the same background from a smoothed, a linear, or a constant model, and the fine detail is then anchored to that background rather than to the starting model.
The final error is governed by the data-resolvable band, matching the near-invariance measured on BP 2004 in Table~\ref{tbl:robust}, where C2F-IFWI stays near $260$~m/s from all three starts while the all-resolution baseline, imprinting initialization-dependent error at the first iteration, degrades from $339$ to $512$~m/s.

\subsection{Design of the activation schedule}
\label{subsec:disc_schedule}
The schedule has two components and two hyperparameters.
The activation index of Eq.~\eqref{eq:m_tau} sets the rate at which levels are admitted.
It is governed by the number of coarse levels $K_{0}$ kept active throughout and by the fraction $\rho$ of training over which the remaining levels are unfrozen.
The per-level weight of Eq.~\eqref{eq:m_wk} sets how each level is introduced once its turn arrives.
We use a cosine ramp in Eq.~\eqref{eq:m_wk} so that a level fades in with zero slope at both ends. This admits the new capacity smoothly rather than as a step. Because the weight multiplies the level's features, it also gates the gradient, so the ill-conditioned gradient of a freshly admitted level is scaled up gradually. In effect, each level receives a short warm-up.

We study each design choice on Overthrust with the settings of Table~\ref{tbl:configs}, varying one factor at a time.

\subsubsection{Ramp fraction $\rho$}
We first vary the fraction $\rho$ over which the fine levels are unfrozen, with the linear time profile fixed.
The ramp-fraction block of Table~\ref{tbl:abl_schedule}(a) reports the final model RMSE.
The result is flat across $\rho \in [0.4, 0.8]$, with a spread below $1$~m/s.

\begin{table}[htbp]
\centering

\renewcommand{\arraystretch}{1.15}
\setlength{\tabcolsep}{10pt}
\begin{minipage}[t]{0.32\textwidth}
\centering
\textit{(a) Ramp fraction} $\rho$ {\footnotesize(linear)}\\[3pt]
\begin{tabular}{cc}
\toprule
$\rho$ & RMSE \\
\midrule
$0.4$ & $196.3$ \\
$0.5$ & $196.4$ \\
$0.6$ & $\mathbf{196.3}$ \\
$0.7$ & $196.2$ \\
$0.8$ & $195.8$ \\
\bottomrule
\end{tabular}
\end{minipage}
\hfill
\begin{minipage}[t]{0.32\textwidth}
\centering
\textit{(b) Time profile} {\footnotesize($\rho=0.6$)}\\[3pt]
\begin{tabular}{lc}
\toprule
Profile & RMSE  \\
\midrule
Linear      & $\mathbf{196.3}$ \\
Cosine      & $195.3$ \\
Square root & $200.5$ \\
Square      & $197.3$ \\
Exponential & $196.9$ \\
\bottomrule
\end{tabular}
\end{minipage}
\hfill
\begin{minipage}[t]{0.32\textwidth}
\centering
\textit{(c) Fixed vs learned}\\[3pt]
\begin{tabular}{lc}
\toprule
Schedule & RMSE \\
\midrule
Fixed (ours) & $\mathbf{196.3}$ \\
Learned      & $204.1$ \\
\bottomrule
\end{tabular}
\end{minipage}
\caption{Schedule ablations on the Overthrust model RMSE. In each panel one factor is varied with the others held at the adopted setting. Bold marks the adopted setting. RMSE in m/s.}
\label{tbl:abl_schedule}
\end{table}

The precise duration of the ramp therefore does not matter, only that the fine levels are deferred.

\subsubsection{Shape of the time profile}
We next fix $\rho=0.6$ and change how the activation index of Eq.~\eqref{eq:m_tau} advances with the iterations.
We replace the linear profile with profiles that open the fine levels earlier (square root), later (square and exponential), or with a cosine ease-in.
The time-profile block of Table~\ref{tbl:abl_schedule}(b) reports the result.
All profiles lie within about $5$~m/s, so the shape is a second-order factor.
The only clearly worse choice is the square-root profile, which activates the fine levels earlier.
This again shows that early activation, not the exact fade, is what reduces accuracy.
We adopt the linear profile for simplicity, since it is within run-to-run variation of the best.

\subsubsection{Fixed versus learned schedule}
Finally, we ask whether the schedule should be learned rather than fixed.
We make the opening times trainable and optimize them jointly on the data misfit.
The last block of Table~\ref{tbl:abl_schedule}(c) compares the two.
The learned schedule converges to a faster opening and a higher final error.
The benefit of deferral is a property of the training dynamics, namely the avoidance of poorly conditioned early updates.
It is not reflected in the instantaneous data misfit, so a gradient on that misfit always favors opening the levels sooner.
Learning the schedule would therefore require a bilevel objective that scores the whole optimization trajectory rather than the current residual, which we leave to future work.

\section{Conclusions}
\label{section: Conclusion}
We presented C2F-IFWI, an implicit full waveform inversion framework whose velocity model is parameterized by a multi-resolution hash coordinate network with a coarse-to-fine activation of the resolution levels.
Rather than activating all hash levels from the start, the schedule keeps only the coarsest levels active and unfreezes the remaining levels from coarse to fine, in step with a low-to-high frequency continuation, so the model is reconstructed from low to high wavenumber.
A frozen level contributes nothing and has no gradient updates until it is activated, which withholds the fine levels until their frequencies are present in the data, stabilizes their optimization, and suppresses cycle skipping.
On the synthetic Overthrust and BP 2004 benchmarks, the coarse-to-fine activation is more accurate than the all-resolution hash baseline at identical parameter count, far less sensitive to the initial model, and robust to strong data noise, and on the Viking field data it produces a clean, interpretable, well-tied image where unconstrained grid FWI degrades.
Future work includes extending the coarse-to-fine representation to elastic and multi-parameter inversion and to three dimensions, and combining it with generative priors for field-scale regularization.

\bibliographystyle{seg}
\bibliography{refs}

\end{document}